\documentclass[sn-mathphys,Numbered]{sn-jnl}

\usepackage{graphicx}%
\usepackage{multirow}%
\usepackage{amsmath,amssymb,amsfonts}%
\usepackage{amsthm}%
\usepackage{mathrsfs}%
\usepackage[title]{appendix}%
\usepackage{xcolor}%
\usepackage{textcomp}%
\usepackage{manyfoot}%
\usepackage{booktabs}%
\usepackage{algorithm}%
\usepackage{algorithmicx}%
\usepackage{algpseudocode}%
\usepackage{listings}%

\usepackage{soul}

\begin{document}

\newcommand\green[1]{\textcolor{green}{#1}} 
\setstcolor{green} 

\title[Article Title]{Patching-based Deep Learning model for the Inpainting of Bragg Coherent Diffraction patterns affected by detectors' gaps}


\author[1,2,*]{\fnm{Matteo} \sur{Masto}}
\equalcont{These authors contributed equally to this work.}

\author[1,2]{\fnm{Vincent} \sur{Favre-Nicolin}}
\author[1]{\fnm{Steven} \sur{Leake}}
\author[1]{\fnm{Tobias} \sur{Schülli}}
\author[3]{\fnm{Marie-Ingrid} \sur{Richard}}
\author[3,*]{\fnm{Ewen} \sur{Bellec}} \email{matteo.masto@esrf.fr, ewen.bellec@esrf.fr}
\equalcont{These authors contributed equally to this work.}

\affil[1]{\orgdiv{ESRF}, \orgname{The European Synchrotron}, \orgaddress{\street{Av. de Martyrs, 71}, \city{Grenoble},\country{France}}}

\affil[2]{\orgname{Univ. Grenoble - Alpes}, \orgaddress{\city{Grenoble}, \country{France}}}

\affil[3]{\orgname{Univ. Grenoble Alpes, CEA Grenoble, IRIG, MEM, NRS}, \orgaddress{\street{17 rue des Martyrs}, \city{Grenoble}, \country{France}}}


\abstract{We propose a deep learning algorithm for the inpainting of Bragg Coherent Diffraction Imaging (BCDI) patterns affected by detector gaps. These regions of missing intensity can compromise the accuracy of reconstruction algorithms, inducing artifacts in the final result. It is thus desirable to restore the intensity in these regions in order to ensure more reliable reconstructions. The key aspect of our method lies in the choice of training the neural network with cropped sections of both experimental diffraction data and simulated data and subsequently patching the predictions generated by the model along the gap, thus completing the full diffraction peak. 
This provides us with more experimental training data and allows for a faster model training due to the limited size, while the neural network can be applied to arbitrarily larger BCDI datasets. Moreover, our method not only broadens the scope of application but also ensures the preservation of data integrity and reliability in the face of challenging experimental conditions.}

\keywords{Convolutional Neural Networks, Inpainting, Bragg Coherent Diffraction Imaging}



\maketitle

\section{Introduction}\label{sec1}



Coherent Diffraction Imaging (CDI) is a lens-less technique that exploits scattering of a coherent X-ray beam to study nano-particles with high spatial resolution \cite{miao_oversampling_2000,miao_approach_2001, Pfeifer2006}.
The resulting diffraction pattern contains information about the three-dimensional (3D) electron density distribution in the material. However, since the phase information is lost during the measurement, iterative algorithms are needed to reconstruct the real-space object. This procedure, referred to as phase retrieval (PR), normally entails alternated projections between direct and reciprocal space, and the application of constraints in both spaces such that the algorithm converges towards the solution \cite{Fienup:78, Fienup:86, Favre-Nicolin2010AnalysisImaging,Marchesini2007ARetrieval,Miao2012CoherentImaging}.

In case of crystalline samples, Bragg Coherent Diffraction Imaging (BCDI) allows to measure the internal strain of particles ranging in size from few micrometers down to 20 nm \cite{Robinson2009, Richard:te5091, Robinson2009CoherentNanoscale, Hofmann20173DNano-crystals}. BCDI enables to observe the strain evolution \textit{in situ} during gas reactions \cite{Carnis2021, Dupraz2022}, in electrochemical environment \cite{Hua2019StructuralOxides, Atlan2023} or to observe the particle modifications over temperature variations \cite{Chatelier:2023}. These reactions often occur at the surface of the nano-particles, thus a good spatial resolution is required in order to follow their evolution by monitoring the effects at the particles' surface. Since the measured intensity ($\mathbf{I}$) corresponds to the squared modulus of the object Fourier Transform (FT), the real-space resolution is inversely proportional to the extent of the recorded diffraction pattern. Consequently, there is a requirement for detectors with large sensing areas to achieve high resolution. \cite{Bond1958OnSignals}.

Standard photon counting detectors are usually assembled out of pixelated chips separated by insensitive gaps. These gaps consist of a few pixel-wide lines, whose size varies according to the detector model. For example, the photon-counting MAXIPIX detector contains a cross-shaped six pixel-wide gap \cite{Ponchut_2011} while the Eiger2M detector \cite{Johnson_2014}, having a larger sensing area, has both 12 and 38 pixel-wide gaps. Technological solutions are on the horizon, \textit{e.g.} the PIMEGA or through silicon \textit{via} technology \cite{Campanelli_2023}, but the majority of pixel detectors available at the time of writing have gaps between active areas.

The acquisition of the 3D BCDI pattern is obtained by rotating the sample and by stacking each 2D detected image for each rotation angle. This implies that the gap \textit{lines} in each 2D image turn into gap \textit{planes} of empty pixels in the full 3D pattern.


The effect of these regions of missing intensity on the recorded diffraction is the corruption of the PR algorithms that eventually leads to the presence of artifacts in the reconstructed real-space object \cite{Carnis2019TowardsReconstructions} (see Supplementary Figure 1). These artefacts become more significant as the BCDI 3D array is large, thus severely limiting the reconstructed object resolution.

Here, we propose to preprocess the 3D experimental BCDI data affected by these gaps using a Deep Learning (DL) inpainting method. Our model is able to make consistent predictions of the in-gap intensity on experimental BCDI data, thus reducing artifacts in the reconstructed object.

Image inpainting has been widely explored in the field of photography and image processing for the restoration of damaged pictures. Many techniques have been developed, from classical polynomial interpolations to more advanced techniques such as diffusion-based methods or sparse representation methods \cite{Jam2021AMethods}. 
More recently image inpainting has been addressed with the use of deep convolutional neural networks which have shown promising and accurate results in different fields \cite{Xiang2023DeepSurvey}. 

In the field of Coherent Diffraction Imaging, and more specifically BCDI, deep learning methods have been exploited for defect identification, classification \cite{Lim2021ADiffraction, Judge2022DefectAI}, and for phase retrieval \cite{Cherukara2018Real-timeNetworks, Chan2021RapidLearning, Yao2022AutoPhaseNN:Imaging}. Image inpainting using DL for X-ray diffraction imaging has already been successfully tested by Bellisario \textit{et al.} \cite{Bellisario:cw5034} on 2D simulated data and by Chavez \textit{et al.} \cite{Chavez:jl5040} on 2D X-ray scattering images. However, a recurrent problem in DL for BCDI is the lack of a large experimental dataset, thus the need to train the model using mostly simulated diffraction data. This limitation often biases the DL models that eventually yield poor results on experimental data. Moreover, these DL models use a fixed input-output size which is inconvenient for practical use since typical experimental BCDI data are cropped and centered during preprocessing, leading to possible different array size each time.

Here, we propose a solution that solves both the limited experimental dataset as well as the size constraint issues for the case of detector gap inpainting through the implementation of a “patching model” trained on small 32$\times$32$\times$32 pixel-size cropped portions ($\mathbf{P}$) of the diffraction patterns. This patching technique allows to use a large number of small portions from experimental BCDI data along with simulated ones. Henceforward, we can use a much lighter and rapidly trainable DL network. Our model can then be applied on large BCDI 3D array, regardless of the data size. The size of $\mathbf{P}$ was chosen such as it was larger than the usual gap size and the finite size oscillations of the intensity pattern.



\section{Results}\label{sec2}
\subsection{Dataset preparation}

The dataset used to train the model contains a mix of simulated and experimental Bragg Coherent Diffraction patterns. Experimental data (ED) were taken from measurements performed at the ID01 beamline of The European Synchrotron (ESRF, in Grenoble, France) \cite{Leake2019TheStructure}. Simulated data (SD) have been constructed following the procedure described in Ref. \cite{Lim2021ADiffraction}, \textit{i.e.} creating 3D face-centered-cubic (FCC) crystals from random atomic elements, crystal shapes and sizes, followed by an energy relaxation using LAMMPS \cite{plimpton_fast_1995}. The 128$\times$128$\times$128 pixel-size BCDI diffraction pattern of the (200) Bragg peak was then calculated using the PyNX package \cite{Favre-Nicolin2020PyNX:Operators} with random particle orientation as well as reciprocal space step size, \textit{i.e.} different oversampling ratio. However, the resulting SD are still very different compared to what was measured experimentally and could bias our DL model, diminishing its applicability to experimental data. To prevent this, we modified the SD by introducing noise in both reciprocal and real space as detailed in Supplementary section S2.

From each entire diffraction pattern, 10 portions $\mathbf{P}$ of 32$\times$32$\times$32 pixel-size have been cropped out pseudo-randomly, see Fig. \ref{fig:process}. Having noticed poorer accuracy for the prediction around low intensity regions, portions from peripheral areas have been preferentially selected over the center of the peak. Thus the final dataset, composed of 50\% ED and 50\% SD, contains 30,000 of these small portions. 

\subsection{Data preprocessing}

\begin{figure}[h!]
\centering
  \includegraphics[width= .75\textwidth]{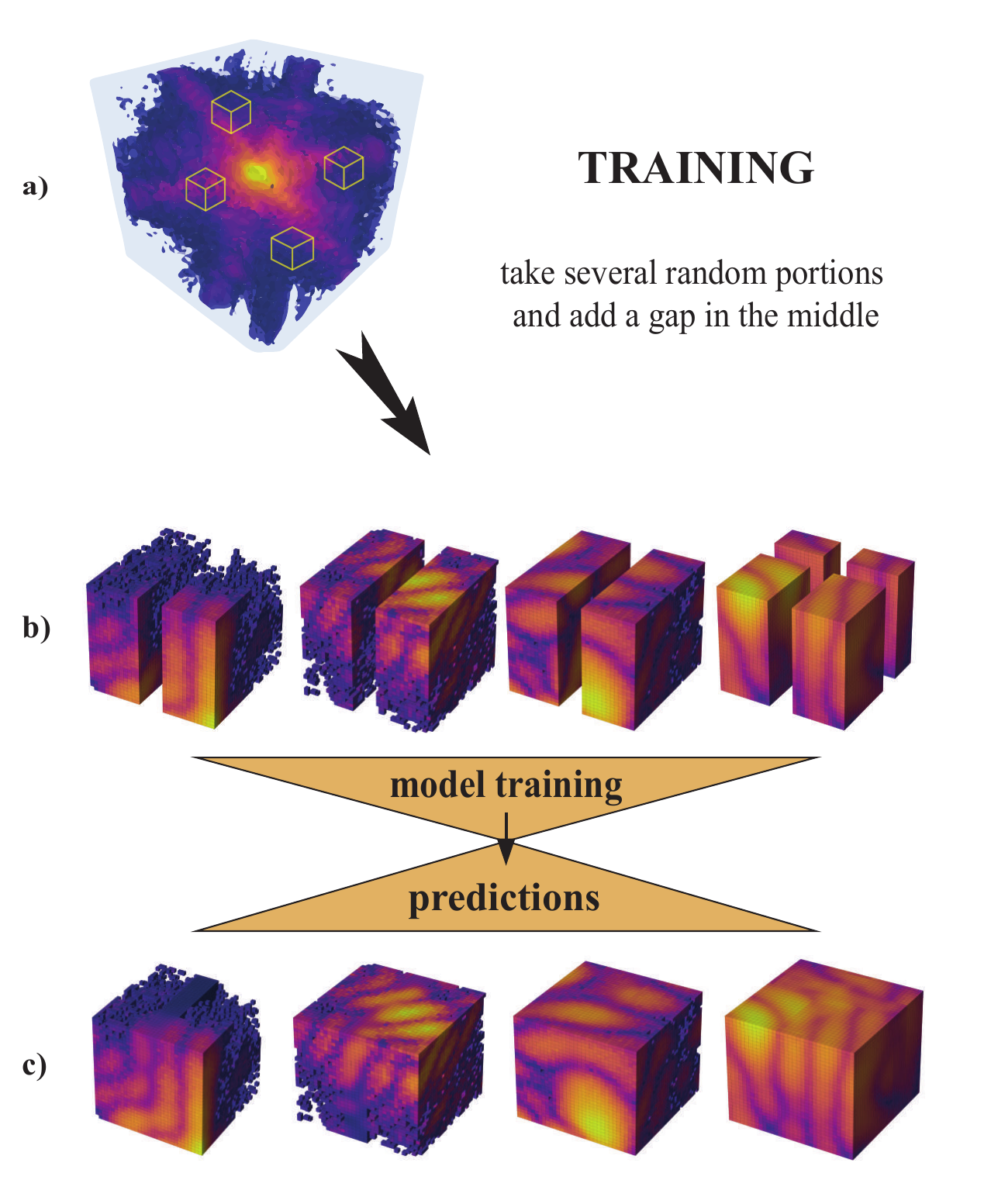}
      \caption{\textbf{Sketch of the data processing and DL model training}. \textbf{a} Large BCDI data where small 32$\times$32$\times$32 pixel-size portions are randomly selected. \textbf{b} Small portions in renormalized log scale and artificially masked with zeros to simulate detector's gaps and used as input of the DL model. \textbf{c} DL model predictions for the corresponding masked inputs with the inpainted gap.}

\label{fig:process}
\end{figure}

During the DL model training, an artificial vertical mask of zero intensity pixels, and of fixed size, was added in the middle of each single portion $\mathbf{P}$, as defined above, in order to simulate the presence of the detector gap (Fig. \ref{fig:process}b). To include the case of cross-shaped gaps, an additional mask, of equivalent size, was applied horizontally at a random position to a certain subset of the training data (see the last example in Fig. \ref{fig:process}b). 

The last preprocessing step transforms the data to a logarithmic scale and normalises each image between 0 and 1 in order to avoid overfitting high-intensity regions over the low-intensity ones. The masked images were then used as model input while the ground truth unmasked images were used in the calculation of the loss function, as a comparison with the DL predictions.

\subsection{Network architecture and training}

\begin{figure}
\includegraphics[trim={2cm 2cm 2cm 2cm}, width=\textwidth]{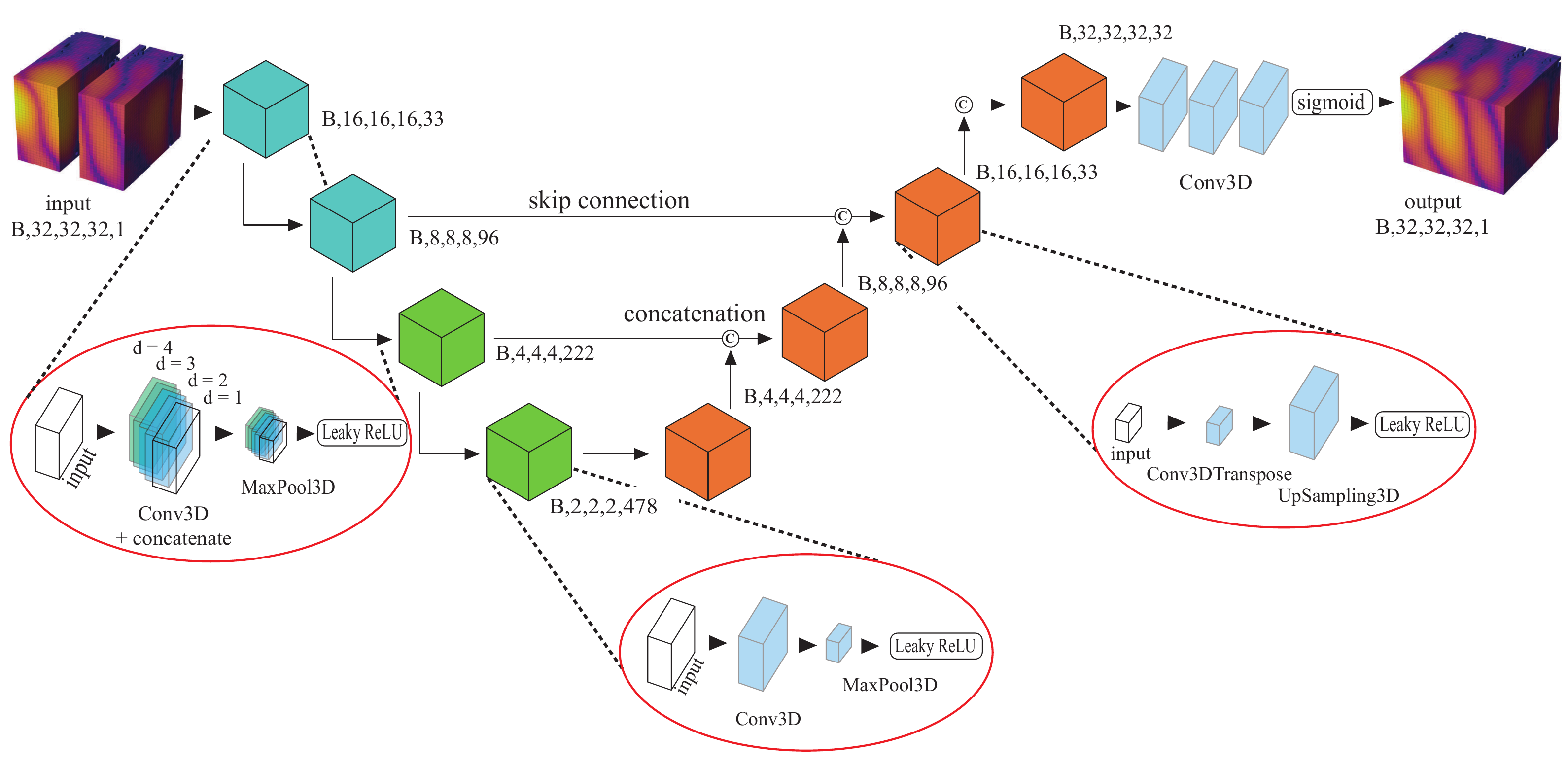}
\caption{\textbf{Architecture of the deep learning model.} A modified U-Net architecture with the use of dilated convolutions in the first two encoder blocks (left red circle) where the input was concatenated to its convolutions with different dilation rates ($d$ parameter) before the MaxPooling layer. The gap-affected small portions $\mathbf{P}$, in batches (B) of 32, were used as input (top left) and progressively sent through the encoder section down to a small volume of 2x2x2 pixels. Each building block of the decoder section (orange cubes) takes as input the concatenation of its previous block's output with the corresponding output of the encoder block of the the same size. The final output (top right) being a batch of inpainted $\mathbf{P}$.}
\label{fig:architecture}
\end{figure}

The adopted DL model was based on the U-Net architecture \cite{Ronneberger2015U-Net:Segmentation}, see Figure \ref{fig:architecture}. It consists of two main blocks, namely the encoder and the decoder. The encoder is composed of 4 convolutional layers followed by MaxPooling and leads to a reduction of the data array size from 32$\times$32$\times$32 down to 2$\times$2$\times$2. The decoder section uses other convolutional and UpSampling layers to enlarge the array back to its original size. Information is transferred between each encoder and decoder layers through \textit{skip connections}, ensuring an easier search for the loss function's absolute minimum \cite{Li2017VisualizingNets}. All encoder and decoder layers use the Leaky ReLU activation function with a slope of 0.2 for negative inputs. Finally, 3 additional convolutional layers are added after the decoder as a way to avoid image smoothing by the array expansion in the decoder. The sigmoid activation function is used as last layer in order to bound the output values between 0 and 1, for the model outputs to have the same intensity range as the inputs.
To improve the receptive field of the first convolutional layers and thus provide higher long-range correlation understanding in the feature extraction, dilated convolutions have been employed with variable dilation rate \cite{Chen2017RethinkingSegmentation}. More precisely, as depicted in Fig.  \ref{fig:architecture}, in the first two encoder blocks the input was concatenated with four different convolutions of itself, each one with a different dilation rate (the \textit{d} parameter in Fig.\ref{fig:architecture}). 
Standard convolutional layers were used in last two encoder blocks as the size of the inputs was already small enough to be treated with normal convolutional layers and a 3$\times$3$\times$3 pixel-size kernel. \\

The network was built using the Tensorflow python library \cite{abadi2016tensorflow} and was trained for 100 epochs, with ADAM optimizer \cite{kingma2017adam} starting with a learning rate of 10$^{-3}$ decreasing progressively using the \emph{ReduceLROnPlateau} callback available in Tensorflow. The shuffled 30,000 small portions dataset was split into a training (93.5\%), validation (4\%) and test (2.5\%) sets. Batches of 32 images were used and training and validation losses were monitored at each epoch in order to avoid overfitting to the training dataset. Inpainted output and ground truth regions were compared using a custom loss function consisting in the sum of three main terms, namely: a Mean Absolute Error (MAE), a Structural Similarity Index perceptual loss \cite{Wang2004ImageSimilarity}, and a MAE on the image gradients.  \\ 

To assess the DL model performances we used the Pearson Correlation Coefficient (PCC). This coefficient measures the linear correlation between two images and in our specific case yields an estimation of the similarity between the DL prediction and the corresponding ground truth, thus an indication of the predictive accuracy of the model. It is defined by the following:




\begin{equation}
PCC = \frac{\sum_{i\in \text{gap}}(\textbf{P}_i^{\text{true}} - \langle \textbf{P}^{\text{true}}\rangle)(\textbf{P}_i^{\text{pred}}-\langle\textbf{P}^{\text{pred}}\rangle)}{\sqrt{\sum_{i\in \text{gap}}^{}(\textbf{P}_i^{\text{true}} - \langle \textbf{P}^{\text{true}}\rangle)^2}\sqrt{\sum_{i\in \text{gap}}^{}(\textbf{P}_i^{\text{pred}}-\langle\textbf{P}^{\text{pred}}\rangle)^2}},
    \label{eq:accuracy}
\end{equation}\\

where $\mathbf{P}^{\text{true}}$ was the 32$\times$32$\times$32 pixel-size ground truth portion in log scale without gap and $\mathbf{P}^{\text{pred}}$ was the same portion where the gap region was inpainted using our DL model. The $\langle \rangle$ symbol corresponds to the average over the gap.  We note that the PCC for identical images was equal to 1.

Table \ref{table:accuracy} shows the average PCC values over a batch of 1000 samples of small portions from experimental BCDI data (where gaps have been artificially added). Vertical gaps of different sizes are considered. As expected, the accuracy decreases when the gap size increases since the prediction of the in-gap fringes becomes more and more difficult. Examples of DL predictions on small BCDI regions with a 6 pixels gap are shown in Supplementary Figs. 7 and 8 for respectively simulated and experimental data, demonstrating accurate in-gap intensity prediction. 

\begin{table}[h]
\caption{\textbf{Average DL model accuracy on 32$\times$32$\times$32 pixel-size small BCDI portions over a batch of 1000 samples}. The accuracy decreases as the gap size increases.}\label{table:accuracy}%
\begin{tabular}{@{}lllll@{}}
\toprule
Gap size (pixels) & \textbf{3}  & \textbf{6} & \textbf{9} & \textbf{12}\\
\hline \\
Pearson Correlation Coefficient  &0.989       & 0.977        & 0.955      & 0.946 \\

\botrule
\end{tabular}
\end{table}

\subsection{ Results in reciprocal space }

In order to make a prediction of the in-gap intensity of a large 3D BCDI array of arbitrary size, we use a ``patching" method. A 32$\times$32$\times$32 pixel-size portion $\mathbf{P}$ centered around the gap of the large image was used as the DL model input and the in-gap intensity was predicted in this region. $\mathbf{P}$ was then repeatedly shifted by 1 pixel at a time along the gap and the prediction was calculated again, until the whole gap intensity was reconstructed. The final step involves averaging the overlapping predicted pixels, which contributes to robust DL predictions even when applied to experimental data. This averaging process helps mitigate potential prediction errors by smoothing them out.

However, this method can be time consuming as a prediction for a cross-shaped gap on 128$\times$128$\times$128 pixel-size BCDI data can take up to 12 minutes. To speed up this process, one can shift $\mathbf{P}$ by more than 1 pixel at a time, drastically decreasing the prediction time down to 1m30s for 4 skipped pixels, without significantly worsening the accuracy. More details on this skip method are available in Section 5 of the Supplementary information.

\begin{figure}[H]
\centering
  \includegraphics[trim={0 0 0 0},width  = \textwidth]{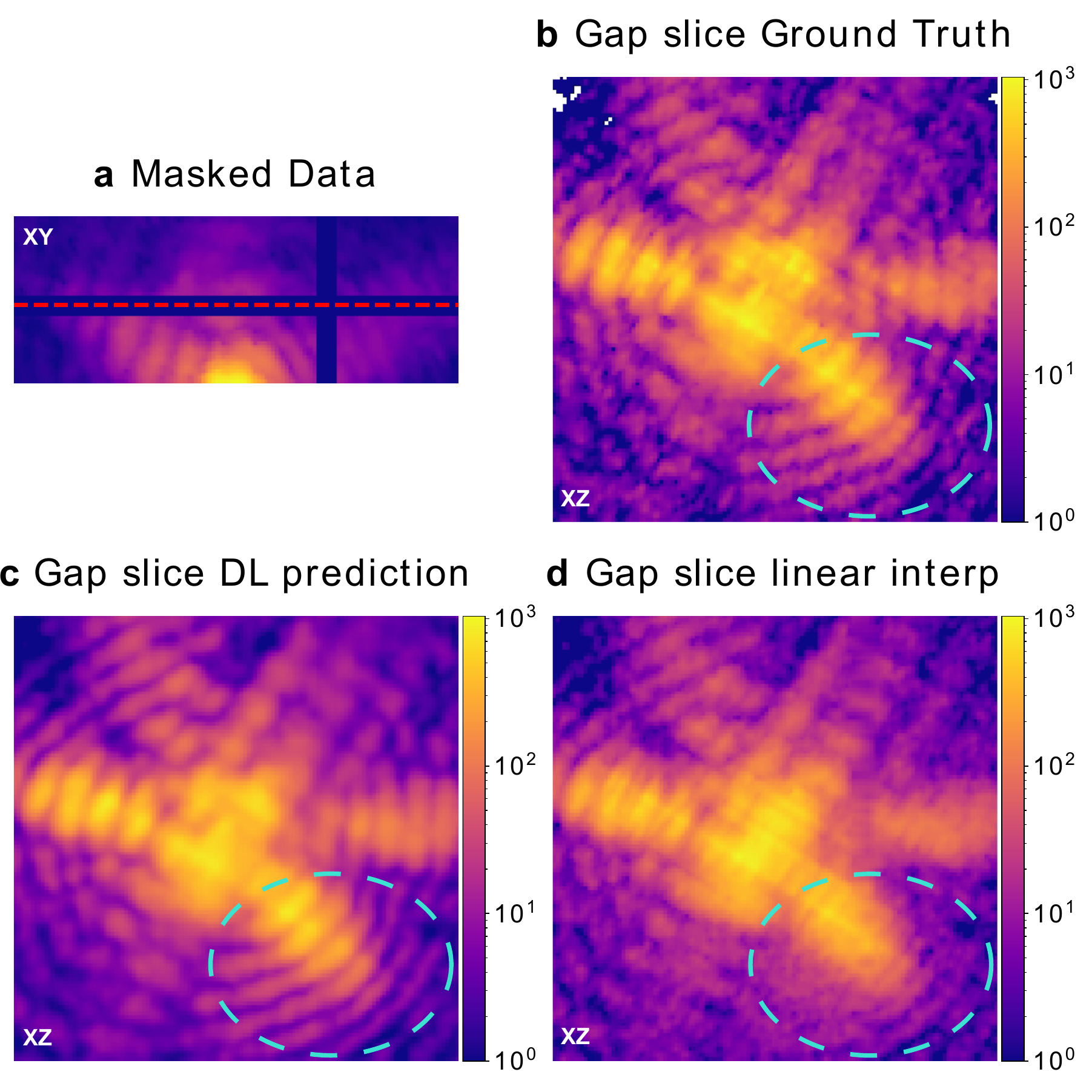}
  \caption{\textbf{\textit{In-gap slice} comparison between DL prediction and standard interpolation on experimental BCDI data.} \textbf{a} 3D experimental BCDI data masked with a cross-like 6 pixel-wide gap - the red dotted line shows the location of the perpendicular `in-gap slices' shown in \textbf{b, c, d}. In-gap slice \textbf{b} ground truth, \textbf{c} DL model prediction, \textbf{d} standard linear-interpolation using pixels immediately around the gap. Only our DL model was able to restore the correct fringes pattern as highlighted by the turquoise dashed circles.}

\label{fig:results}
\end{figure}

In order to test the DL model accuracy on large BCDI data, we use the patching method on a large experimental BCDI array where a handmade 6 pixel-wide cross-shaped gap was added. Figure \ref{fig:results}\textbf{a} displays the position of the gap in the XY plane while Fig. \ref{fig:results}\textbf{b} shows the ground-truth in-gap intensity in the XZ plane. Our DL model prediction is shown in Fig. \ref{fig:results}\textbf{c}, where the fringes pattern was accurately reproduced. One can notice that the "grainy" features of the ground-truth are not reproduced due to the intrinsic denoising effect induced by the model training process \cite{krull2019noise2void}.

As a comparison, a standard linear interpolation (LI) is shown in Fig. \ref{fig:results}\textbf{d}. The cubic and nearest-neighbour interpolations are illustrated in Supplementary Fig. 9. 

 An improvement of the result from the DL model with respect to standard interpolation algorithms was observed in all the cases, in particular when comparing the fringes pattern in the bottom left of Figs. \ref{fig:results}\textbf{c} and \textbf{d}. Where LI fails to reproduce the oscillations, DL succeeds. This was expected, since standard interpolations algorithms have no \textit{a priori} knowledge of the oscillatory nature of diffraction fringes.

\begin{figure}[H]
\centering
  \includegraphics[trim={3cm 3cm 3cm 3cm},width  = \textwidth]{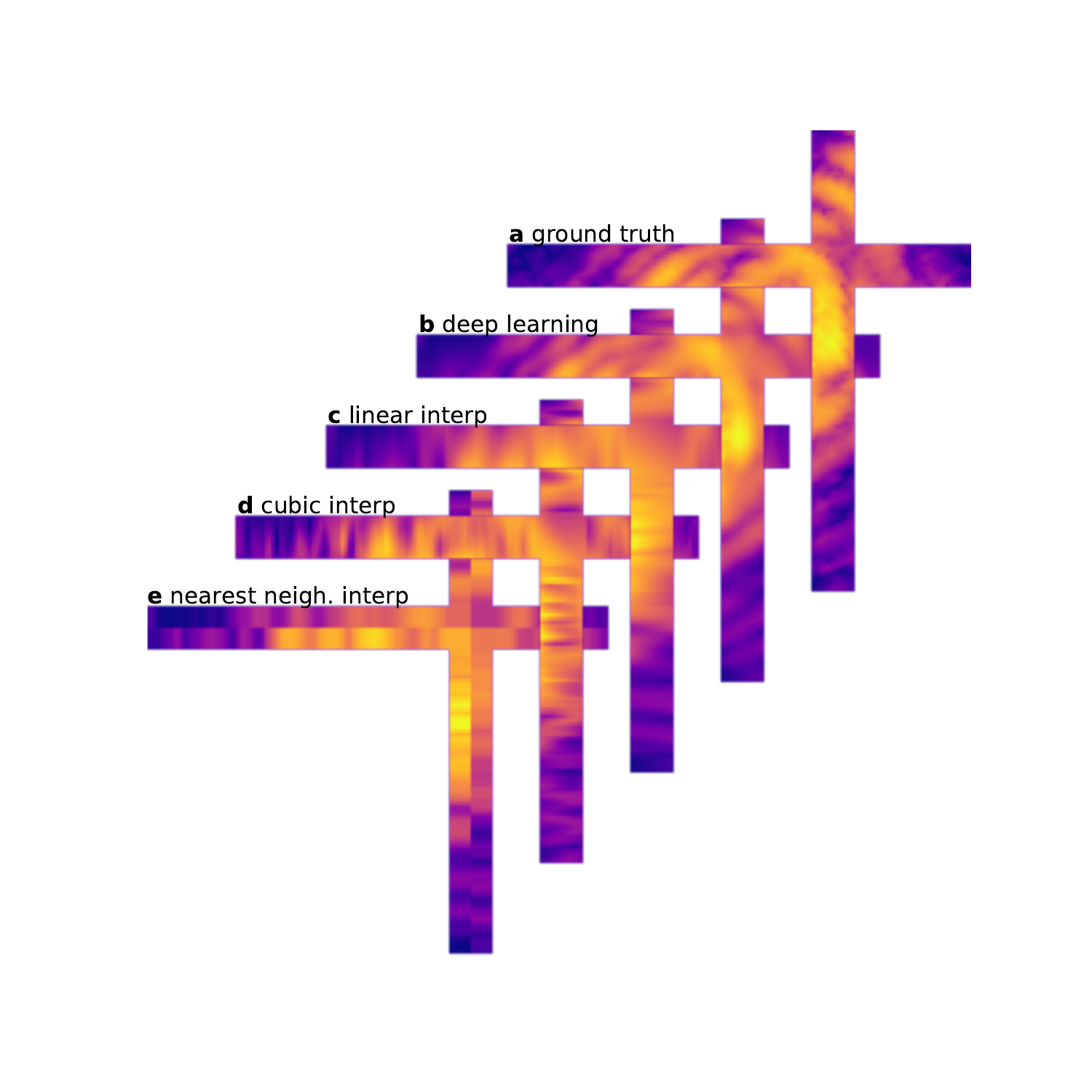}
  \caption{\textbf{Comparison between DL prediction and several standard interpolations across the gap.}
   \textbf{a} Ground truth intensity in the 12 pixel-wide cross-shaped gap (only the gap area is shown). \textbf{b} DL prediction. \textbf{c, d, e} respectively linear, cubic and nearest-neighbour interpolation. Only the DL model was able to recover the accurate fringes curvature across the gap.}

\label{fig:results12px}
\end{figure}

This is further demonstrated in Fig. \ref{fig:results12px} where the in-gap intensity is shown in the XY plane for a gap size of 12 pixels. The DL algorithm (Fig. \ref{fig:results12px}\textbf{b}) was able to predict the correct fringes curvature across the gap. On the other hand, the standard interpolations (see Figs. \ref{fig:results12px}\textbf{c}-\textbf{d}-\textbf{e}) neglect this curvature and reproduce straight oscillations perpendicular to the edges of the gap region.


\subsection{Performances assessment}

\begin{figure}
\centering
    \includegraphics[width= \textwidth]{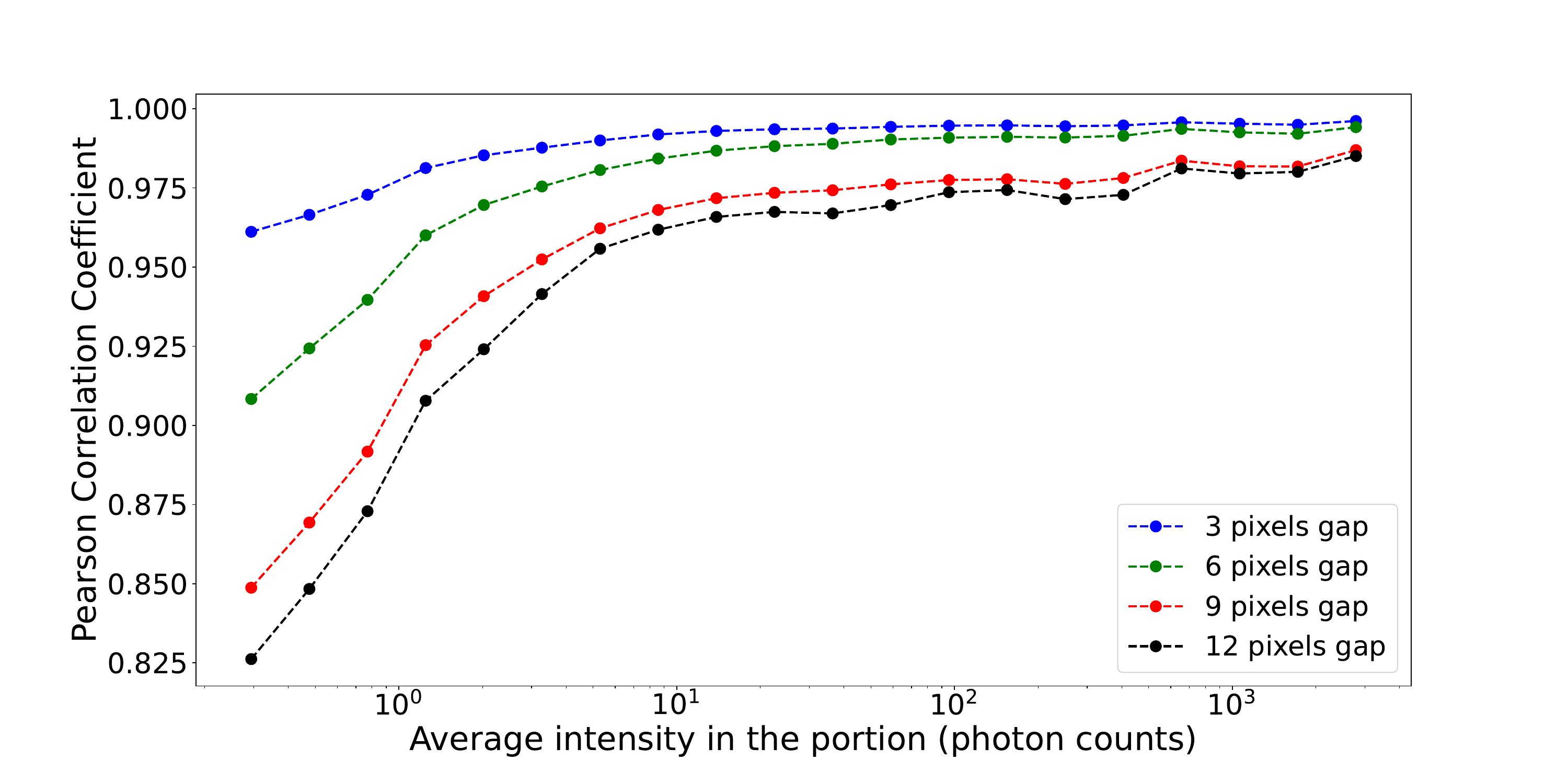}
  \caption{\textbf{Prediction accuracy \textit{vs} average intensity in cropped portion.} The model prediction becomes more accurate as the overall intensity inside the considered portion increases. Conversely, in cases of low photon counts - indicating a prevalence of noise within the portion—the predictions were more prone to inaccuracies.}

\label{fig:acc_int}
\end{figure}

In this section we discuss the accuracy of our model with respect to: (i) the amount of intensity inside the cropped portion $\mathbf{P}$  and (ii) the oversampling ratio. In order to assess the model accuracy for the first case, we used a 128$\times$128$\times$128 pixel-size experimental diffraction pattern and we randomly cropped out 1000 portions $\mathbf{P}$ of 32$\times$32$\times$32 pixels. A vertical gap was placed in the middle of each $\mathbf{P}$ and the DL model was used to predict the in-gap intensity. The PCC accuracy as given in Table \ref{eq:accuracy} was then calculated for each $\mathbf{P}$ individually and its average is shown as function of the average photon count in $\mathbf{P}$ (see Fig. \ref{fig:acc_int}).
Lower PCC scores are obtained when the average intensity in the region is smaller, which is expected as the absence of significant features, that are lost in the Poisson noise, prevents accurate DL prediction. Moreover, as expected, from Fig. \ref{fig:acc_int} emerges that the smaller the gap size, the better the accuracy of the prediction.\\

\begin{figure}
\centering
  \includegraphics[width= \textwidth]{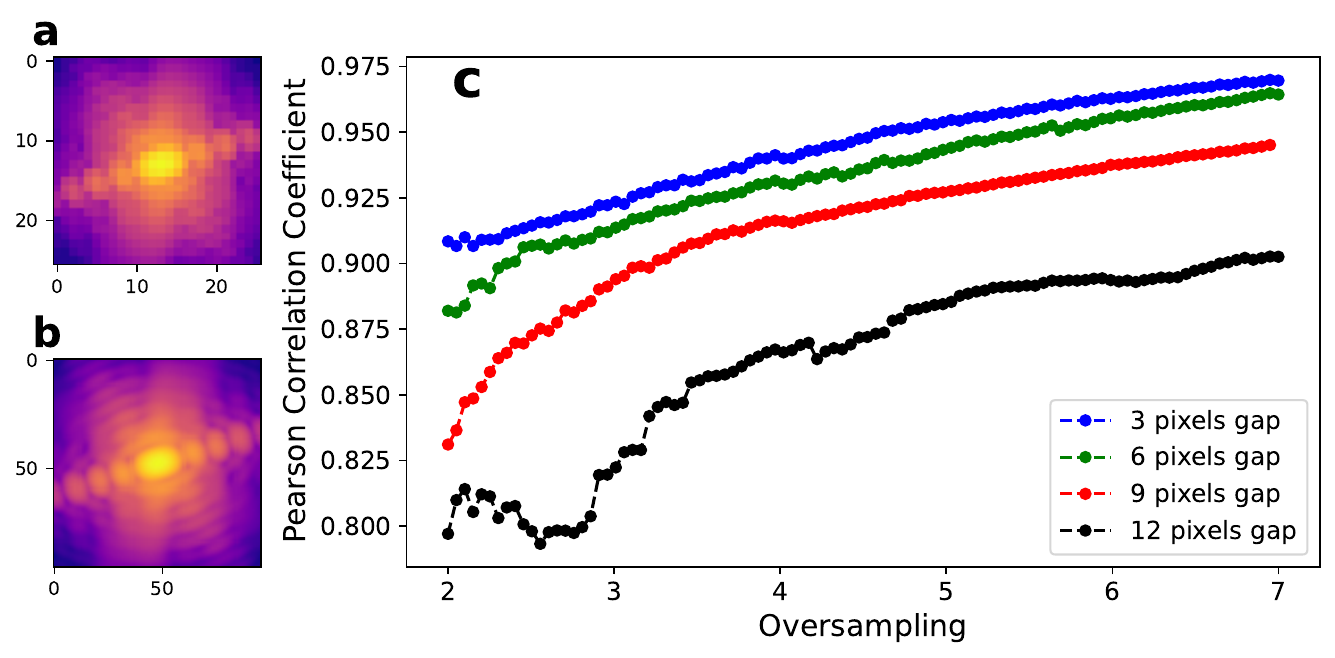}
  \caption{\textbf{Prediction accuracy vs oversampling.} \textbf{a} Low and \textbf{b} high oversampling simulated BCDI array of the same region. \textbf{c} Prediction accuracy as function of BCDI oversampling ratio for different gap size.}

\label{fig:acc_ovs}
\end{figure}

In order to compute the model accuracy as function of the oversampling ratio, we simulate BCDI arrays of the same region for different oversampling ratios (ORs) as shown in Figs. \ref{fig:acc_ovs}
\textbf{a}-\textbf{b}. Since a different OR implies different array size, comparing the model accuracy is not straightforward. To do so, we make the prediction of the full image using the method illustrated in Supplementary Fig. 13. For each OR, a  vertical gap mask was applied to the whole BCDI array and the DL prediction was calculated. The gap was then shifted and this procedure was repeated until the whole BCDI array was predicted using our model, thus leading to a full BCDI predicted image. The PCC shown in Fig. \ref{fig:acc_ovs}\textbf{c} was then calculated using the whole BCDI array for different ORs and model gap sizes. As expected, the predictions are more accurate for large ORs and small gap sizes (\textit{i.e.}, large oscillation periods relative to the gap width). 
Some prediction examples are given in Supplementary Fig. 14.


\subsection{Reconstructions in real-space} 
\subsubsection{Simulated real-space result}

\begin{figure}[H]
\centering
  \includegraphics[width= .8\textwidth]{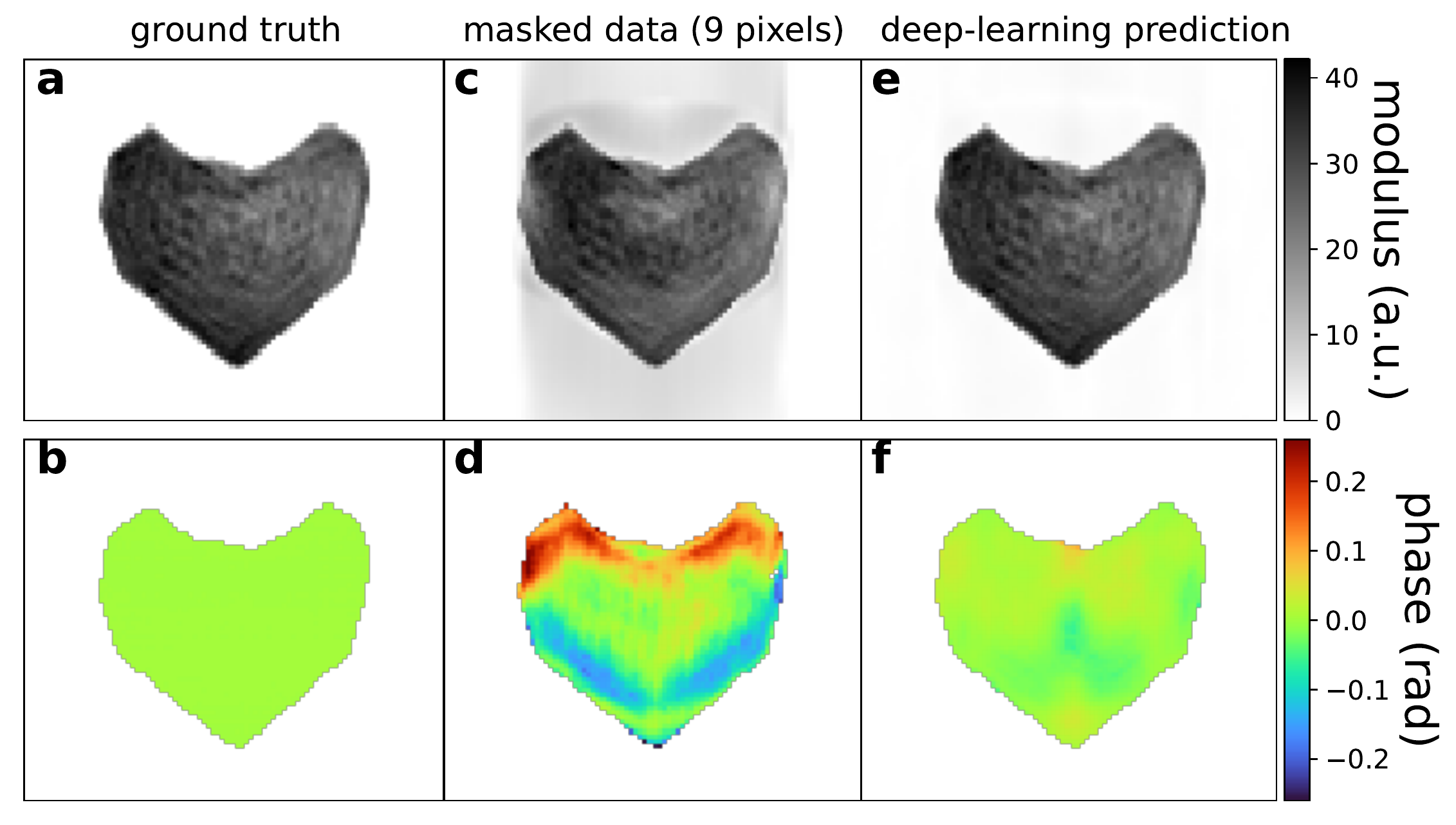}
  \caption{\textbf{DL inpainting result on real-space object reconstruction.}
   \textbf{a-b} Central slice of the ground truth object modulus and phase obtained from a simulated diffraction pattern with no gap. \textbf{c-d} Reconstruction with a 9 pixel-wide cross-like gap affected diffraction pattern. \textbf{e-f} Reconstruction after DL gap inpainting, drastically reducing the artefacts induced by the gap. The corresponding diffraction patterns are available in Supplementary Fig. 15. Note that in this example the phase of the ground truth object has been artificially set to zero (contrary to Fig. \ref{fig:reconstructions_exp}) for an easier comparison.}

\label{fig:reconstructions}
\end{figure}

Since the final goal of the BCDI technique, before physical analysis, is the reconstruction of the real-space complex object, we assess here the reconstructed object quality before and after DL gap inpainting. A simulated BCDI array was used starting from a reconstruction by Carnis \textit{et al.} \cite{Carnis2019TowardsReconstructions}. After the reconstruction from experimental diffraction pattern, the real space phase of the particle was artificially set to zero (see Figures \ref{fig:reconstructions}\textbf{a}-\textbf{b}), making the evaluation of the gap effect easier. From this reference ``ground truth" object $\mathbf{O}$, the simulated diffracted amplitude correspond to $\mathbf{A}=\text{FT}\left[\mathbf{O}\right]$, where FT is the Fourier transform.


A 9 pixel-wide cross-shaped gap mask was then applied to $\mathbf{A}$ (see Supplementary Fig. 19\textbf{b}) and the corresponding real space object was calculated from the inverse FT (Figs. \ref{fig:reconstructions}\textbf{c}-\textbf{d}). The presence of the gap in the diffraction pattern induces artefacts in real-space manifesting as non-zero module values outside the support region along the directions perpendicular to the gap planes (Fig. \ref{fig:reconstructions}\textbf{c}).
Most importantly, the gap induces variations in the object phase, and thus the reconstructed displacement field and strain \cite{Godard:te5072}, especially near the sample surfaces (Fig. \ref{fig:reconstructions}\textbf{d}). 
Here, a phase variation of $\pm$0.2 radians is observed in Fig. 7\textbf{d} resulting in an error of $\pm$7pm in the lattice displacement field for the (111) Pt reflection.
These artefacts are particularly problematic in the cases of (electro-)catalytic experiments \cite{Atlan2023} or \textit{in-situ} gas experiments \cite{ulvestad_situ_2016,kim_active_2018,abuin_coherent_2019,kawaguchi_gas-induced_2019,Dupraz2022} where the chemical reactions occur at the nano-particle's surfaces and can be studied by following the strain evolution in these regions. The presence of a large gap, or a gap close to the centre of the Bragg peak, could lead to a physical misinterpretation from a poorly reconstructed phase.


Afterwards, our DL model was used to predict the in-gap masked intensity (see Supplementary Fig. 15\textbf{c}), and the corresponding object was computed \textit{via} the inverse FT using the ground truth reciprocal space phase (Figs. \ref{fig:reconstructions}\textbf{e}-\textbf{f}). The artefacts on the reconstructed modulus disappear almost entirely. Furthermore, the reconstructed phase standard deviation is five times lower than the one calculated for the case with a gap and does not present any large variations close to the surfaces, showing that our DL method is suitable for the object reconstruction.



Referring to the work of Carnis \textit{et al.} in Ref. \cite{Carnis2019TowardsReconstructions}, we evaluate the Root Mean Squared Error (RMSE) values of the strain for the particle in Supplementary Fig. 18 for different gap sizes. The wider the gap, the larger the variation from the mean, hence the less precise is the obtained strain distribution.
However, it is clearly visible from the same figure that the restoration of the diffraction intensity using our DL method significantly reduces the error on the strain calculation. Also note that the \textit{mean} value of the strain obtained from masked diffraction patterns differs from the expected zero, as depicted in Supplementary Fig. 17.



\subsubsection{Experimental real-space result}

\begin{figure}[H]
\centering
  \includegraphics[width= .7\textwidth]{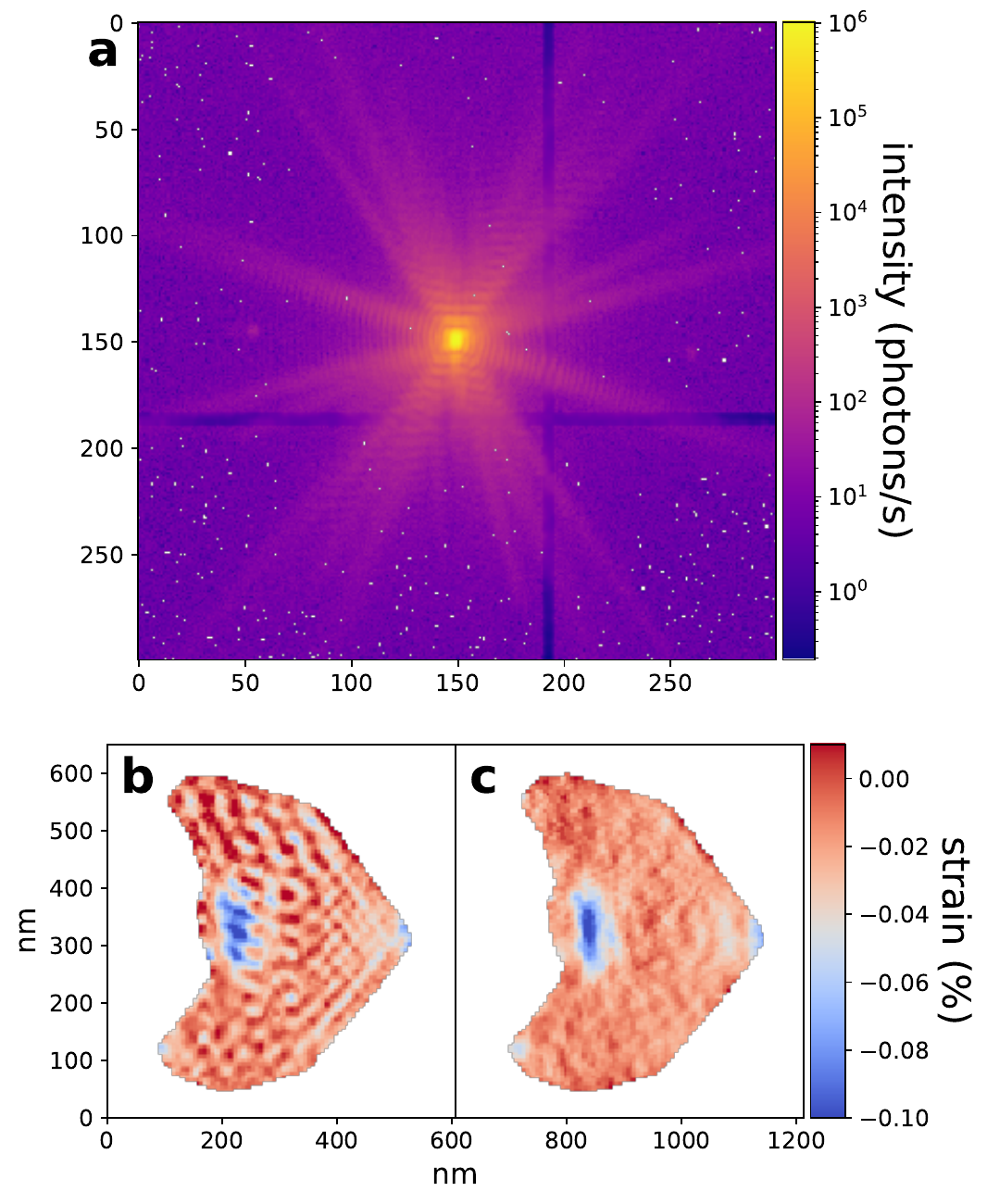}
  \caption{\textbf{DL inpainting result on high-resolution experimental data containing a real gap.} \textbf{a} DL inpainted high-resolution BCDI experimental array containing a real gap. \textbf{b} Object strain reconstruction leaving in-gap pixels free during PR. \textbf{c} Reconstructed strain using DL inpainted gap. The strong oscillation artefacts visible in \textbf{b} are removed by the DL inpainting.}

\label{fig:reconstructions_exp}
\end{figure}

In order to obtain a nano-particle reconstruction with high spatial resolution, one generally has to measure a relatively large BCDI array. With typical photon counting detectors, this leads to a region with a large gap as shown in Supplementary Fig. 1\textbf{a}. One common solution is to run the Phase Retrieval (PR) algorithms leaving the in-gap pixels free.
However, with this approach PR algorithms often overestimate the intensity distribution inside the gap, leading to strong oscillation artefacts in the phase and strain of the reconstructed object as shown in Fig. \ref{fig:reconstructions_exp}\textbf{b}.

On the other hand, by inpainting the gap with our DL model before the PR (Fig.\ref{fig:reconstructions_exp}\textbf{a}), the resulting strain map in Fig. \ref{fig:reconstructions_exp}\textbf{c} does not show any of these artefacts anymore, indicating that the in-gap prediction is accurate. Furthermore, we tested other methods for a comparison, namely: leaving pixels free only at the streaks position and setting the in-gap intensity to 0. The results are depicted in Supplementary Figs. 19-20-21, and show that DL inpainting is the best way to obtain a reliable high-resolution reconstruction from BCDI data with gaps. A second high-resolution example is shown in Supplementary Fig. 22.

\subsection{Fine Tuning}

\begin{figure}[H]
\centering
   \includegraphics[trim={0 0 0 0},width  = \textwidth]{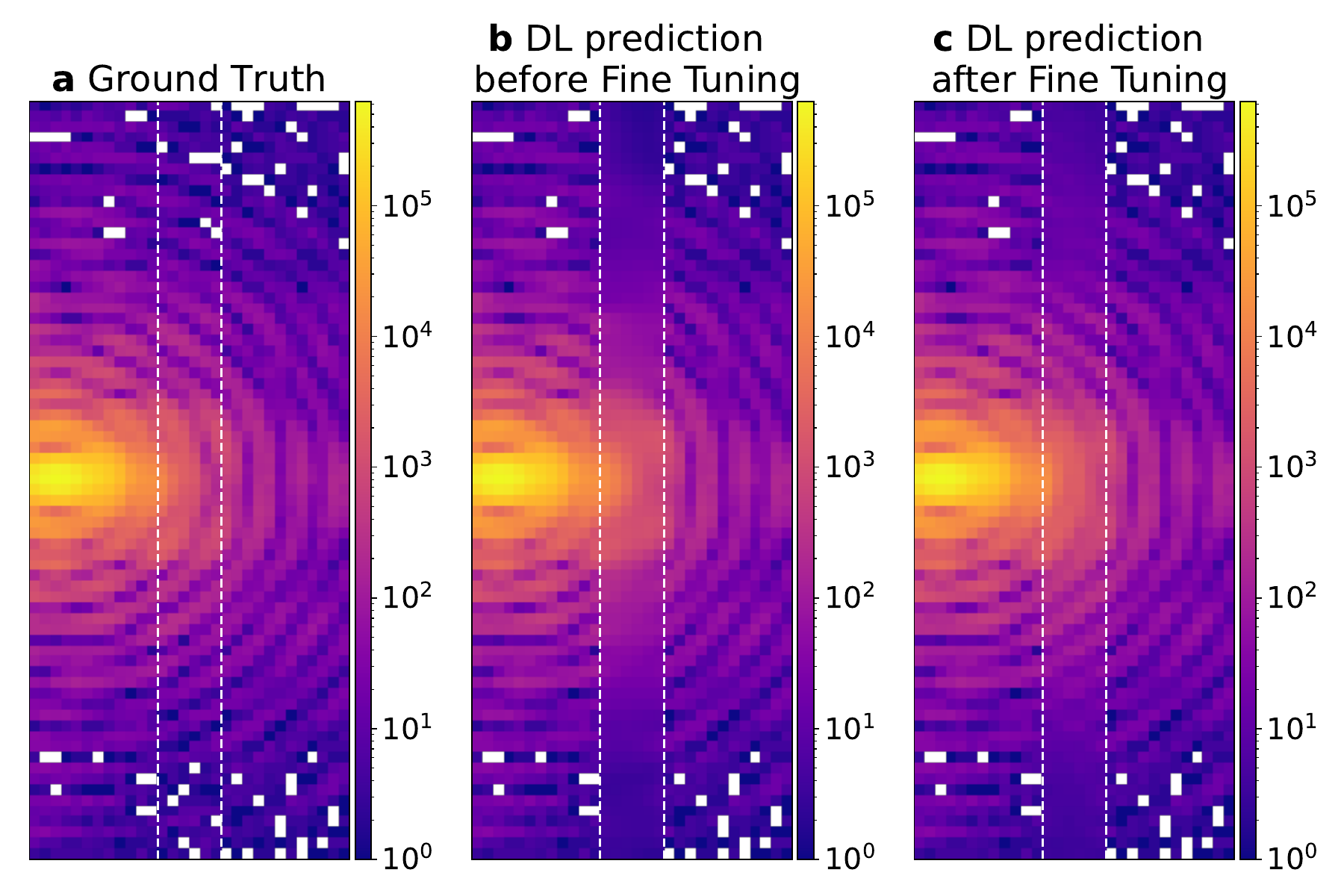}
  \caption{\textbf{Fine Tuning.} \textbf{a} The central slice of an experimental diffraction pattern in log scale. An artificial 6 pixel-wide vertical gap was added in the region between the two white dashed lines. \textbf{b} The corresponding slice after DL inpainting where the fringes pattern is not correctly retrieved. \textbf{c} The same slice of the inpainted image after 2 epochs of fine-tuning of the DL model. The fringes pattern was more reliably recovered.}

\label{fig:finetuning}
\end{figure}

There may be cases in which the DL model does not yield satisfactory predictions inside the gap, such as when the target image is too different from the training dataset, as shown in Fig. \ref{fig:finetuning}\textbf{b}. To overcome these situations, it is possible to fine-tune the model using a specific dataset obtained only from the target image. Our approach involves a secondary short training phase for the model, conducted on a limited dataset (6400 portions) derived from a random sub-sampling of the same 3D diffraction pattern affected by gaps that we aim to restore. This training exclusively uses portions of the detector that remain unaffected by gaps. The second training occurs typically within 2 to 5 epochs and usually takes up to one or two minutes.\\ By performing this fine-tuning, the model is biased on purpose to operate with specific features of the interested image (oversampling, particle shape, detector, \textit{etc}) thus improving the performance on the real gap prediction (see Fig. \ref{fig:finetuning}\textbf{c}). We emphasize that this fine-tuning procedure is only advised when the prediction obtained with the pre-trained model is relatively poor.

\section{Conclusion}

In the present work, a Deep-Learning based approach for inpainting 3D BCDI arrays affected by a detector gap is proposed. The key point of our method is the use of a ``patching" technique where only small portions of the BCDI arrays are used during training. This technique offers several benefits, namely: (i) it effectively removes the constraint on the size of the BCDI array, meaning that there is no need to train different models for different array sizes; (ii) given the small volume of the portions, the training of the model is faster and most importantly (iii) cropping small portions of large data leads to a drastic increase of the amount of experimental data available during the training, removing possible biases that often occur using only a simulated dataset. Our model achieves high accuracy on experimental BCDI data and was able to remove possible reconstruction artefacts on the real-space object, especially in the case of high-resolution BCDI data.

This deep learning ``patching" approach could be applied to other imaging techniques with missing pixel problems and a lack of a large experimental training datasets, such as CDI, ptychography or any other techniques with image spatial correlations.

\backmatter

\bmhead{Supplementary information} 
Supplementary information is available for this paper.

\section*{Declarations}

\begin{itemize}
\item Funding: \\
This project has been partly funded by the European Union’s Horizon 2020 Research and Innovation Programme under the Marie Sklodowska-Curie COFUND scheme with grant agreement No. 101034267 and the European Research Council (ERC) under the European's Horizon 2020 research and innovation programme (grant agreement No. 818823).
\item Availability of data and materials: The datasets used during the current study available from the corresponding author on reasonable request.
\item Code availability: The codes for this study are available and accessible via this link \url{https://github.com/matteomasto/Patching_DL}
\end{itemize}



\bibliography{sn-bibliography}

\end{document}


\title{Supplementary Information: Patching based deep learning model for inpainting of detector gap - affected coherent diffraction images.}

\date{\today}

\maketitle

\section{S1. Gap problem in BCDI}

\begin{figure}[ht!]
\centering
\includegraphics[width=1\textwidth,angle=0]{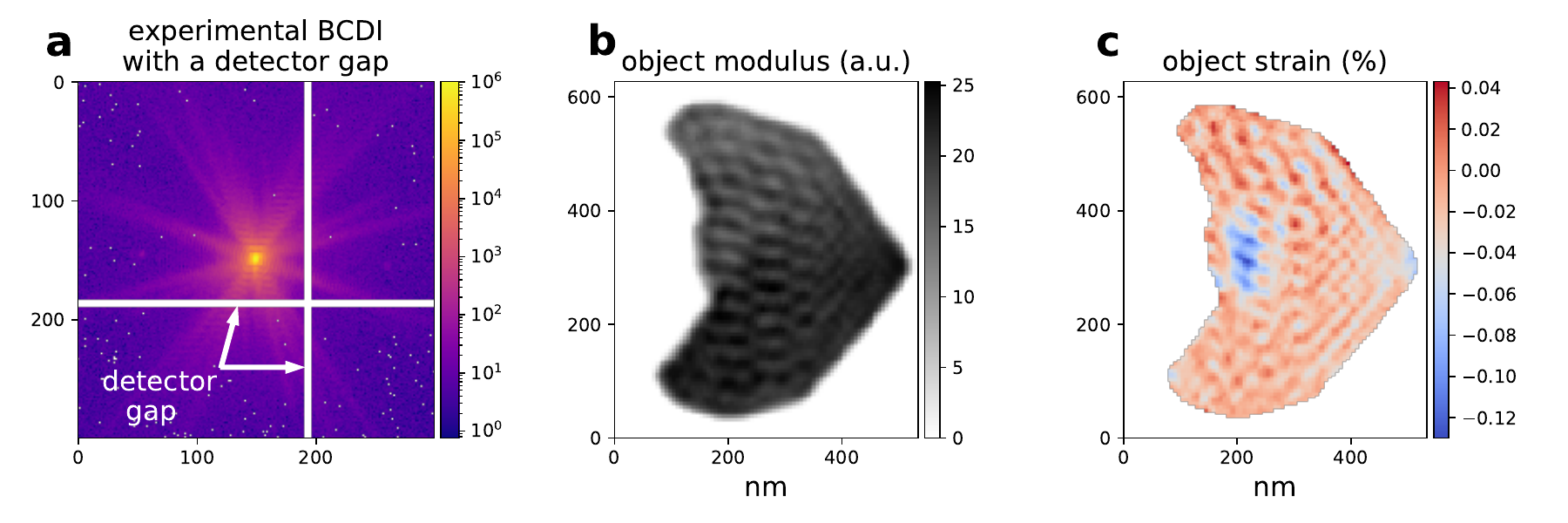}
\caption{\textbf{Gap problem in BCDI.} \textbf{a} BCDI array with gapped region. \textbf{b} Reconstructed object modulus and \textbf{c} strain. Oscillations artefacts induced by the gap presence are visible.}
\label{fig:gap_problem}
\end{figure}

Supplementary Fig. \ref{fig:gap_problem}\textbf{a} shows a typical example of Bragg Coherent Diffraction Imaging (BCDI) data having a detector gap issue. Figures \textbf{b}-\textbf{c} show the reconstructed object modulus and strain respectively. Despite the gap representing only 4\% of the BCDI array and being relatively far from the peak centre, visible oscillation artefacts are observed on the reconstructed object. This gap problem prevents reliable high-resolution BCDI reconstructions.

\section{S2. Simulation of realistic BCDI data}

\subsection{S2.1 Differences between simulation and experimental BCDI data}

Simulated BCDI dataset were created with the same method as in  \textit{Lim et al}\cite{Lim2021ADiffraction}. First, perfect crystal atomic positions are created using MERLIN \cite{Rodney2010}, having different particles shapes (Wulff, Winterbottom, octahedron, cubic) with random set of parameters including particle size, shape, surface ratios and atomic elements. These particles are then relaxed using LAMMPS \cite{PLIMPTON19951} to obtain a realistic strain distribution. The corresponding reciprocal space complex amplitude array $\mathbf{A}$ is then calculated with the kinematic sum using PyNX (see Ref. \cite{Favre-Nicolin2020PyNX:Operators}). Different oversampling ratios\footnote{The oversampling ratio is defined for each dimension as the ratio between the array size and the crystal size in direct space. According to Nyquist theorem, to reconstruct the signal in direct space it is necessary to sample at a frequency at least double the maximum frequency of the signal. In an experiment this means moving the detector along the propagation direction until the resolution on the detector is at least of two pixels per fringe.}  have been considered. The corresponding BCDI array used to train our model is given by $\mathbf{I} = |\mathbf{A}|^2$.

However, the resulting intensities $\mathbf{I}$ are still very different from real experimental data as shown in Supplementary Fig. \ref{fig:simu_diffraction_compare} where both the projection (\textbf{a}) and central slice (\textbf{c}) are compared with the ones of a real experimental BCDI array having a similar oversampling ratio (see Supplementary Fig. \ref{fig:simu_diffraction_compare}\textbf{b} and \textbf{d}). First, the experimental data looks more "roundish" than simulated data (compare \textbf{a} and \textbf{b}). This is due to the fact that our simulated particles have very well-defined facets with sharp corners, which is usually not the case for experimental particles. Smoothing the particles corners can remove this difference. Furthermore, experimental data contains Poisson noise and thus empty pixels far from the central peak (see \textbf{d}) while our simulated data contains no noise and intensity visible everywhere, even far from the center.

\begin{figure}[ht!]
\centering
\includegraphics[width=.7\textwidth,angle=0]{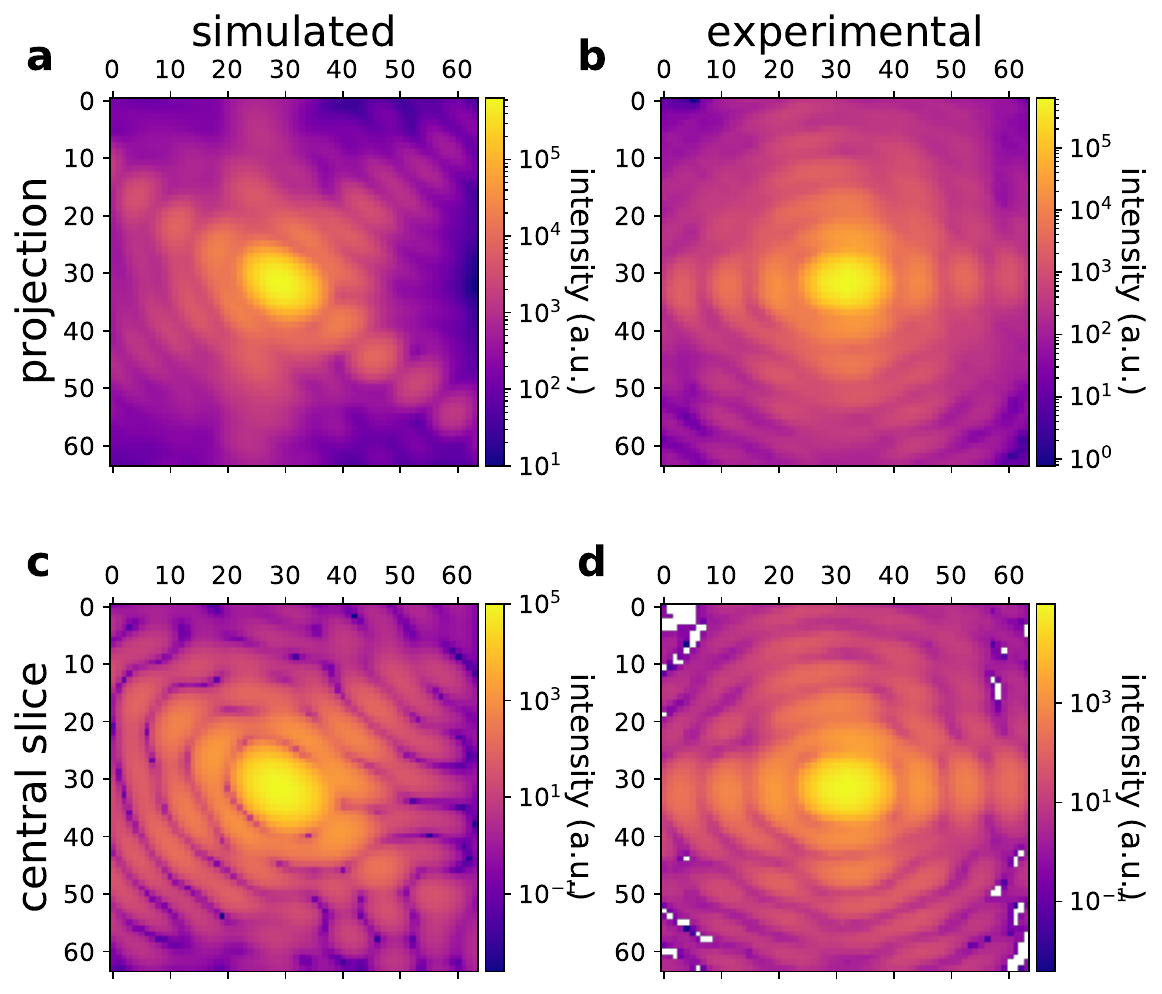}
\caption{\textbf{Comparison of simulated and experimental BCDI data.} \textbf{a}, \textbf{b} projection of simulated and experimental BCDI arrays. \textbf{c}, \textbf{d} corresponding central slice.}
\label{fig:simu_diffraction_compare}
\end{figure}

Additionally, the difference between simulated and experimental data can be visible in real space, when looking at the corresponding complex object $\mathbf{O}$. This object is calculated using the inverse Fourier transform (FT) of the complex BCDI amplitude $\mathbf{O} = \text{FT}^{-1}\left[\mathbf{A}\right]$. The comparison of the object modulus $|\mathbf{O}|$ for simulated and reconstructed experimental data is shown in Supplementary Fig. \ref{fig:simu_module_compare}. We can note that, the simulated object has sharper edges than the experimental one and some line artefacts are visible outside the object in Supplementary Fig. \ref{fig:simu_module_compare}a. Furthermore, the histogram of $|\mathbf{O}|$ is shown in Supplementary Figs. \ref{fig:simu_module_compare}c-d. The histogram's peak sharpness indicates the homogeneity of $|\mathbf{O}|$. Since the full width at half maximum (FWHM) is larger for the experimental peak (0.186) than for the simulated one (0.117), the simulated $|\mathbf{O}|$ is too homogeneous compared to real experimental data. 

\begin{figure}[ht!]
\centering
\includegraphics[width=.7\textwidth,angle=0]{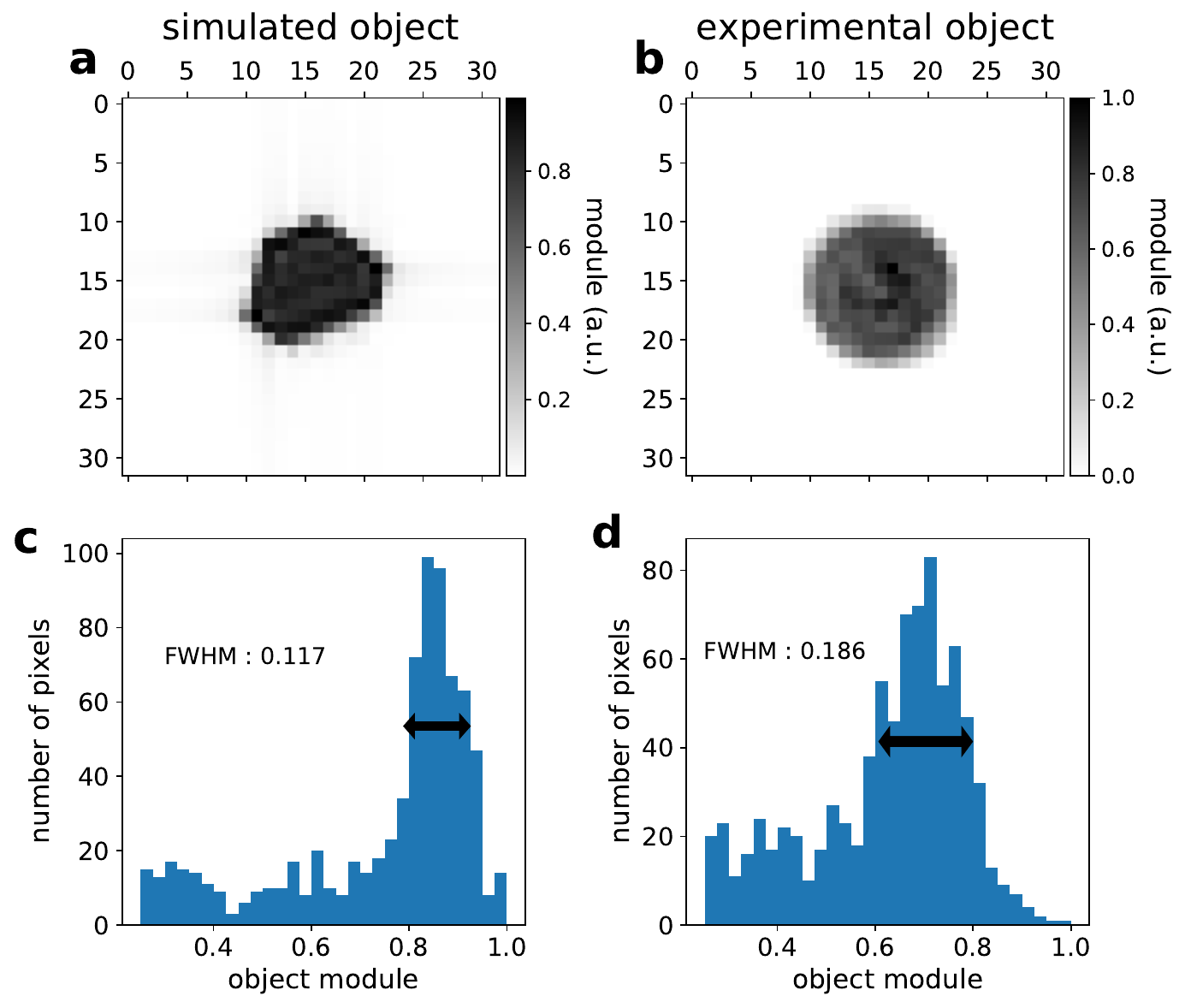}
\caption{\textbf{Comparison between simulated and experimental reconstructed object.} \textbf{a}, \textbf{b} are respectively the central slice of the simulated and experimental object modulus $|\mathbf{O}|$. \textbf{c}, \textbf{d} histogram of $|\mathbf{O}|$ showing that the peak of the experimental object is larger than the simulated one, meaning that the reconstructed experimental object module is less homogeneous than the simulated one.}
\label{fig:simu_module_compare}
\end{figure}

\subsection{S2.2 Real space modifications}

In order to make our simulated data more similar to experimental ones, we first make modifications on the real space object $\mathbf{O}$ as shown in Supplementary Fig. \ref{fig:simu_modif_module}. We use a gaussian convolution to smooth the particle boundaries and corners and remove all non-zero pixels outside a support (from \textbf{b} to \textbf{c}). Secondly, an additional random noise is added in order to remove the homogeneity of the object modulus ($|\mathbf{O}|$). This additional noise is a random Gaussian correlated profile as defined by \textit{Wu et al.}\cite{Wu2021Three-dimensionalNetworks}, where the correlation length is randomly chosen for each simulated data.

\begin{figure}[ht!]
\centering
\includegraphics[width=1\textwidth,angle=0]{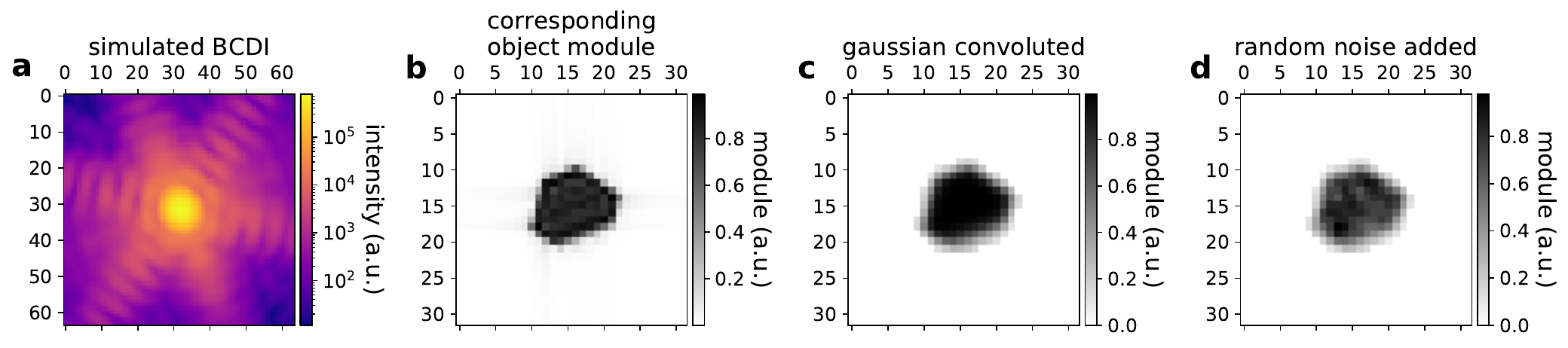}
\caption{\textbf{Simulated object modulus modification.} \textbf{a} Central slice of a simulated BCDI array. \textbf{b} Modulus of the corresponding complex object $|\mathbf{O}|$. \textbf{c} $|\mathbf{O}|$ after a real space Gaussian convolution. \textbf{d} $|\mathbf{O}|$ with an additional random Gaussian correlated noise.}
\label{fig:simu_modif_module}
\end{figure}

Furthermore, the simulated object phase results from the particle relaxation using LAMMPS. In order to increase the dataset diversity and make our deep learning more robust, the phases of some simulated objects are replaced by a random Gaussian correlated profile with random phase amplitude. An example of a complete simulated object modification is shown in Supplementary Fig. \ref{fig:simu_modif_phase}.

\begin{figure}[ht!]
\centering
\includegraphics[width=.7\textwidth,angle=0]{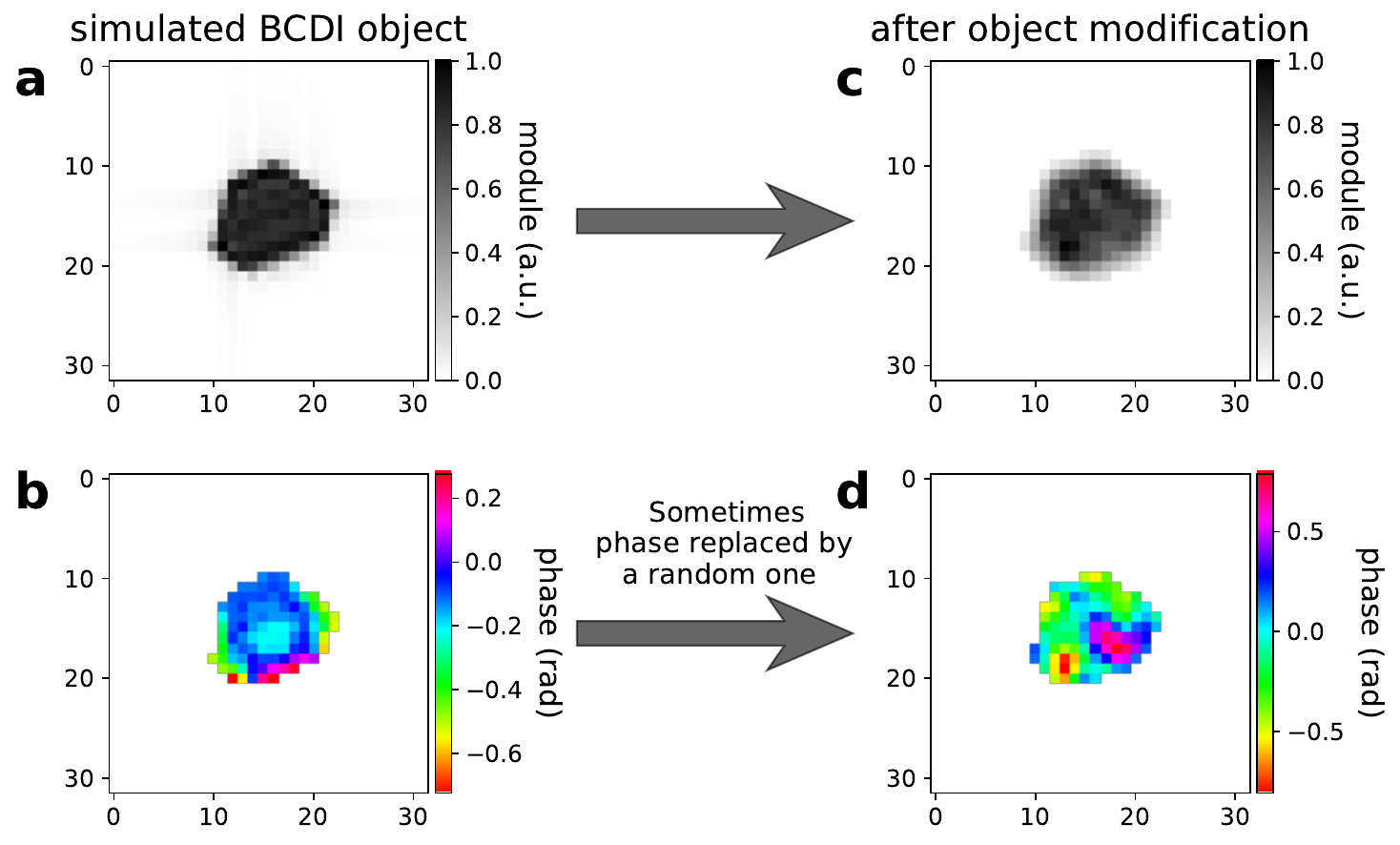}
\caption{\textbf{Random simulated object modification.} \textbf{a}, \textbf{b} Initial simulated BCDI object modulus and phase. \textbf{c} Modulus after Gaussian smoothing and random noise addition. \textbf{d} Some phases of simulated objects are replaced by a random Gaussian correlated profile with random amplitude.}
\label{fig:simu_modif_phase}
\end{figure}

\subsection{S2.3 Reciprocal space modifications}

\begin{figure}[ht!]
\centering
\includegraphics[width=.7\textwidth,angle=0]{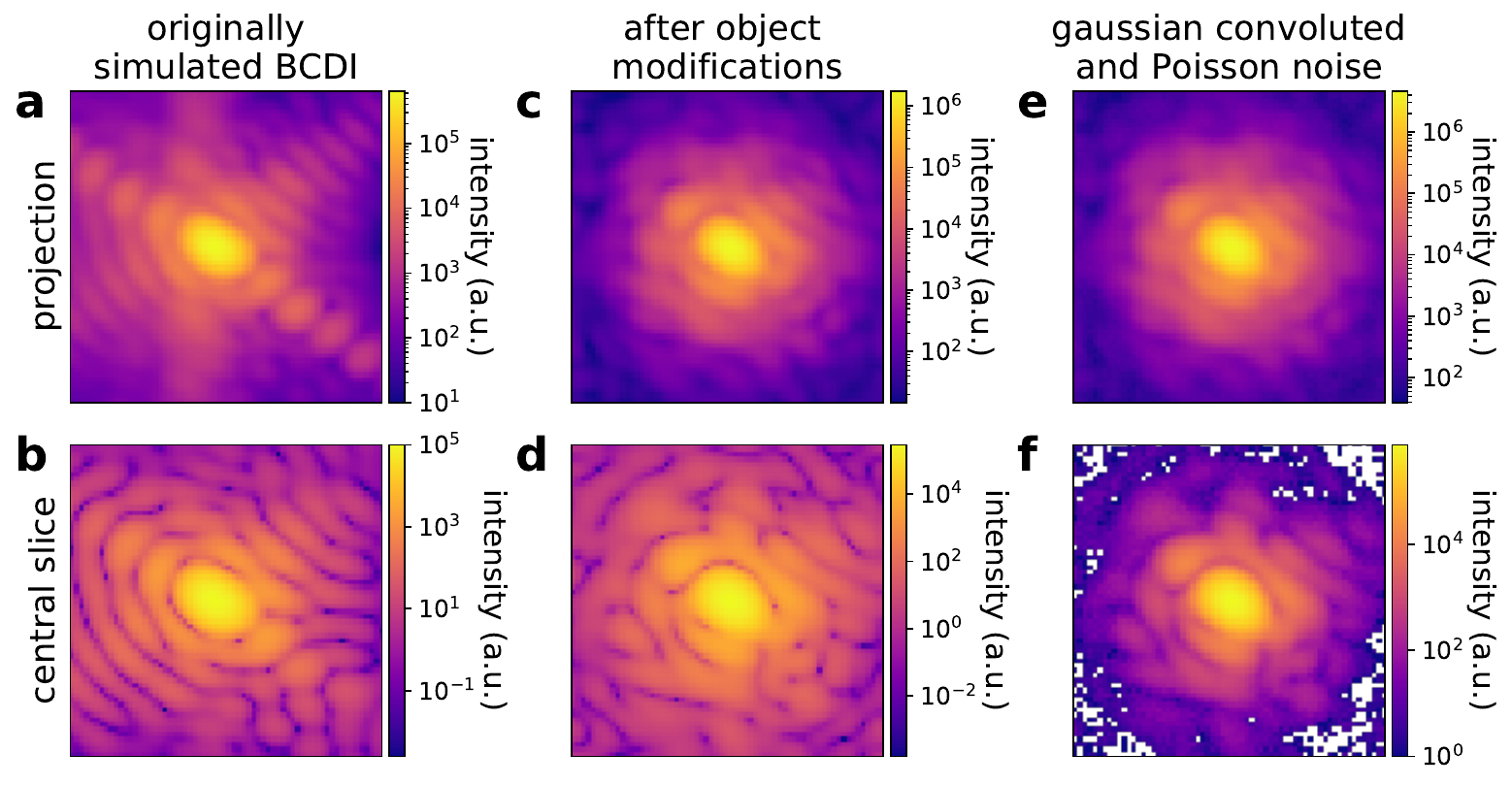}
\caption{\textbf{Random modification of the simulated BCDI intensity.} \textbf{a}, \textbf{b} Projection and central slice of the original simulated BCDI array. \textbf{c}, \textbf{d} Same after the real space object modifications. \textbf{e}, \textbf{f} Same after Gaussian convolution and applying Poisson noise statistic with random amplitude.}
\label{fig:simu_modif_diffraction}
\end{figure}

Finally, after modifying the real space simulated object, we perform some additional corrections on the corresponding BCDI array $\mathbf{I} = |\text{FT}\left[\mathbf{O}\right]|^2$. Supplementary Figs. \ref{fig:simu_modif_diffraction} \textbf{a}-\textbf{b} show the original simulated BCDI array projection and central slice. The same ones are shown after the real space object modification in Supplementary Figs. \ref{fig:simu_modif_diffraction}\textbf{c}-\textbf{d}. Our final adjustments shown in Supplementary Figs. \ref{fig:simu_modif_diffraction}\textbf{e}-\textbf{f} are a Gaussian convolution in order to smooth the Bragg peak fringes and Poisson noise with a random amplitude resulting in empty pixels far from the peak center (as it is the case on experimental data). These randomly modified simulated BCDI arrays are used to train our deep learning model and make it more robust for gap inpainting on experimental data.

\newpage
\section{S3. Small portions prediction}

Our inpainting patching method presented in the main text allows for a smaller and more robust DL model able to yield accurate predictions on experimental data. The model is trained to make predictions on small 32x32x32 portions ($\mathbf{P}$) of both simulated and experimental large BCDI data. Some examples of model prediction on small portions of simulated data are shown in Supplementary Fig. \ref{fig:small_simulated}. The gap is randomly chosen to be either a fixed vertical gap in the middle of the array or a cross-shaped gap with a random vertical position (5$^{th}$ column in Supplementary Fig. \ref{fig:small_simulated} and 3$^{rd}$-5$^{th}$ columns in Supplemenatry Fig. \ref{fig:small_experiment}).

\begin{figure}[ht!]
\centering
\includegraphics[width=1\textwidth,angle=0]{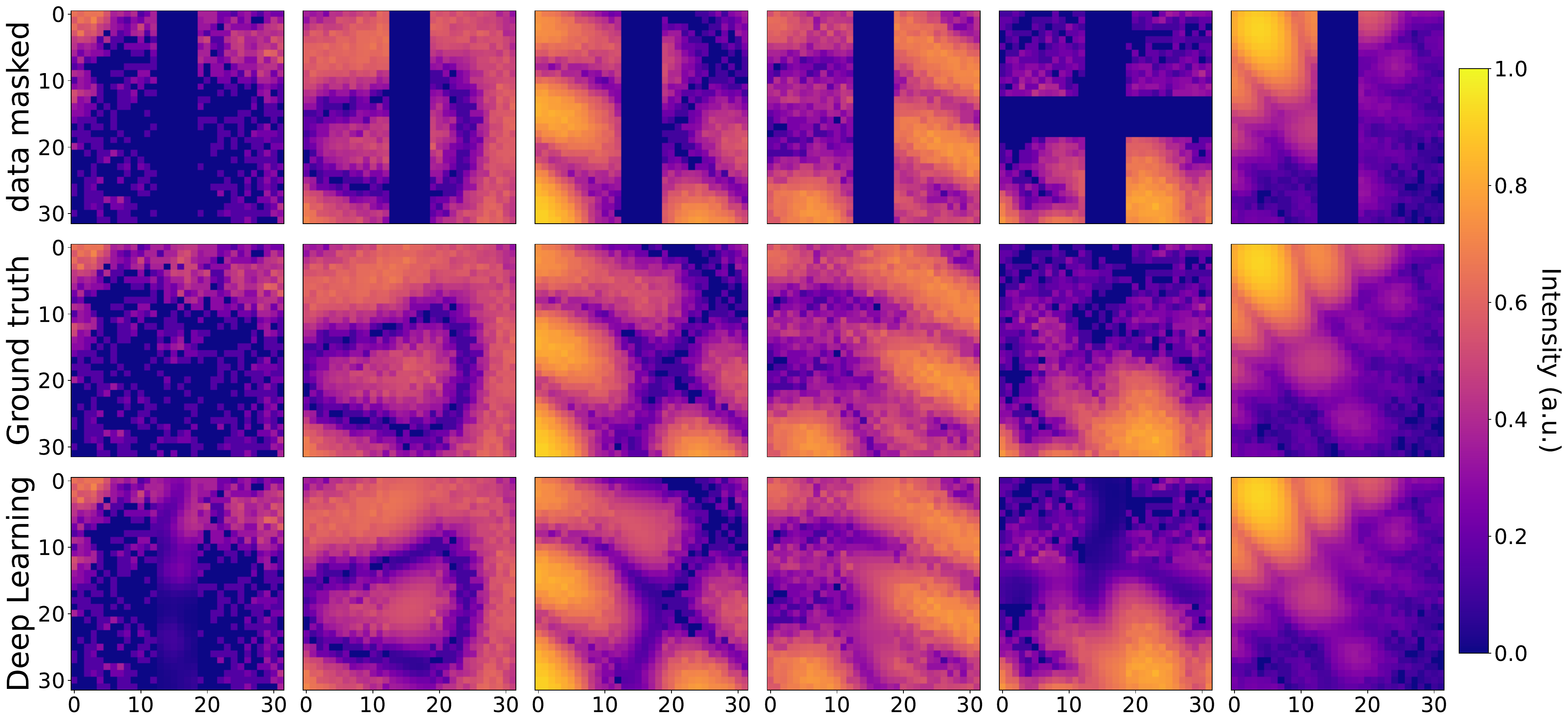}
\caption{\textbf{Model prediction on small 32x32x32 pixel-size portions of simulated BCDI data.} 1$^{st}$ row is the data masked with a random gap. 2$^{nd}$ row is the ground truth and 3$^{rd}$ row is the deep learning model prediction.}
\label{fig:small_simulated}
\end{figure}

Our model predictions on small portions of experimental data acquired at the ID01 beamline of the ESRF-EBS are shown in Supplementary Fig. \ref{fig:small_experiment}. The results are similar to the ones obtained with simulated data, showing that the use of experimental data during training makes the model robust. This eventually allows for efficient predictions on large experimental BCDI data with gap. 

One can note that our model prediction is ``smoother" in low intensity regions compared to the ground truth (see 2$^{nd}$ and 4$^{th}$ columns in Supplementary Fig. \ref{fig:small_simulated}). This is explained by the fact that the ``grainy" aspect of the ground truth in low intensity regions is due to Poisson noise. Since this noise is random, the model can not predict the specific noise configuration inside the gap and thus the best prediction the model can make is the noise average. This phenomenon was already reported in the literature like in the Noise2Void denoising method\cite{krull2019noise2void}.

\begin{figure}[ht!]
\centering
\includegraphics[width=1\textwidth,angle=0]{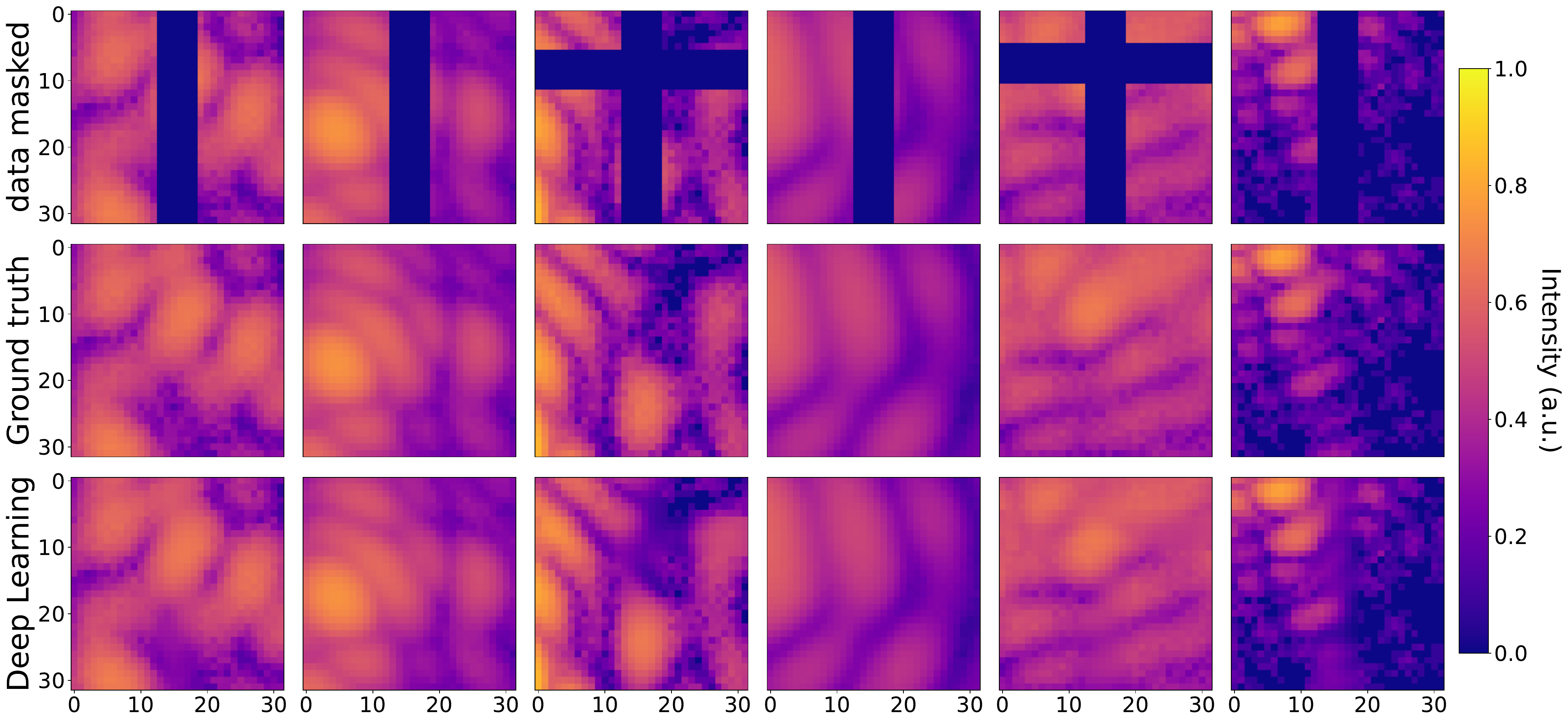}
\caption{\textbf{Model prediction on small 32x32x32 pixel-size portions of experimental BCDI data.} 1$^{st}$ row is the data masked with a random gap. 2$^{nd}$ row is the ground truth and 3$^{rd}$ row is the deep learning model prediction.}
\label{fig:small_experiment}
\end{figure}

\newpage
\section{S4. Comparison with standard interpolation inpainting}

In this section, for the sake of completeness, we show the in-gap predictions provided by nearest-neighbor and cubic interpolations. 

\begin{figure}[ht!]
\centering
\includegraphics[width=0.6\textwidth,angle=0]{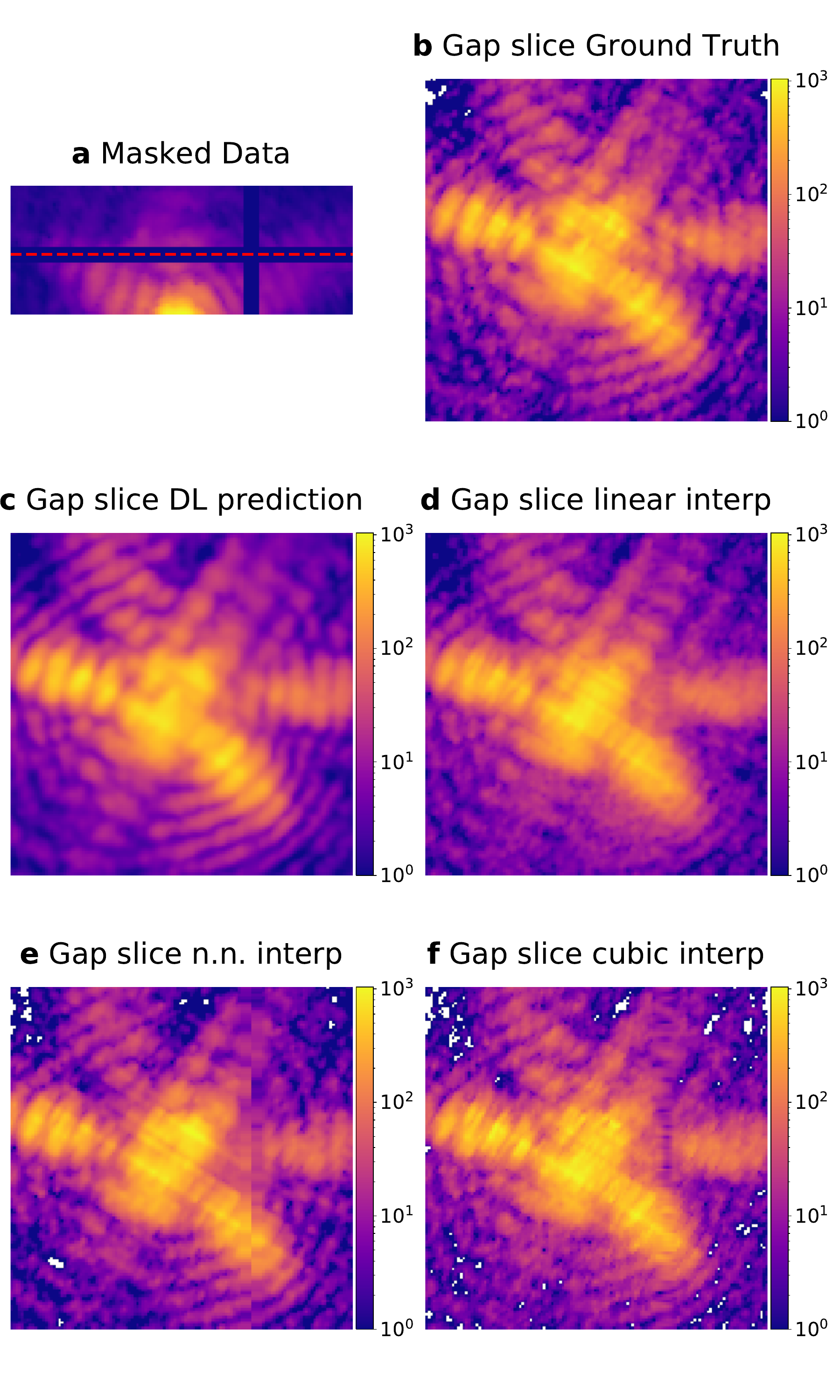}
\caption{\textbf{In-gap slice comparison between DL prediction and standard interpolation on experimental BCDI data}. \textbf{a} Original data masked with a cross-like 6 pixel-wide gap. \textbf{b} Ground truth in-gap slice of the non-masked data along the red dotted line in \textbf{a}. \textbf{c} In-gap slice prediction using DL model, \textbf{d} using linear interpolation, \textbf{e} nearest - neighbor interpolation and \textbf{f} cubic interpolation.}


\label{fig:slice_full}
\end{figure}


\newpage

\section{S5. Skip pixel for large prediction}

Our patching method consists in sliding a 32x32x32 pixel-size portion ($\mathbf{P}$) over the gap area and making a prediction at each position. These predictions are then averaged in overlapping regions. This technique also makes our model more robust when predicting experimental data since potential artefacts are averaged out. This sliding technique has 1 free parameter: the number of ``skipped pixels" $k$ which represents by how much $\mathbf{P}$ is shifted between one small prediction and the next one along the gap. One can choose to shift $\mathbf{P}$ by only one pixel at a time ($k=0$), obtaining the best prediction possible. However, this takes up to 11 minutes for a 128x128x128 pixel-size BCDI data array with a cross-shaped gap. Moving $\mathbf{P}$ by more than 1 pixel at a time, the prediction time rapidly decreases following a power law as shown in Supplementary Fig. \ref{fig:skip_time}.

\begin{figure}[ht]
\centering
\includegraphics[width=.7\textwidth,angle=0]{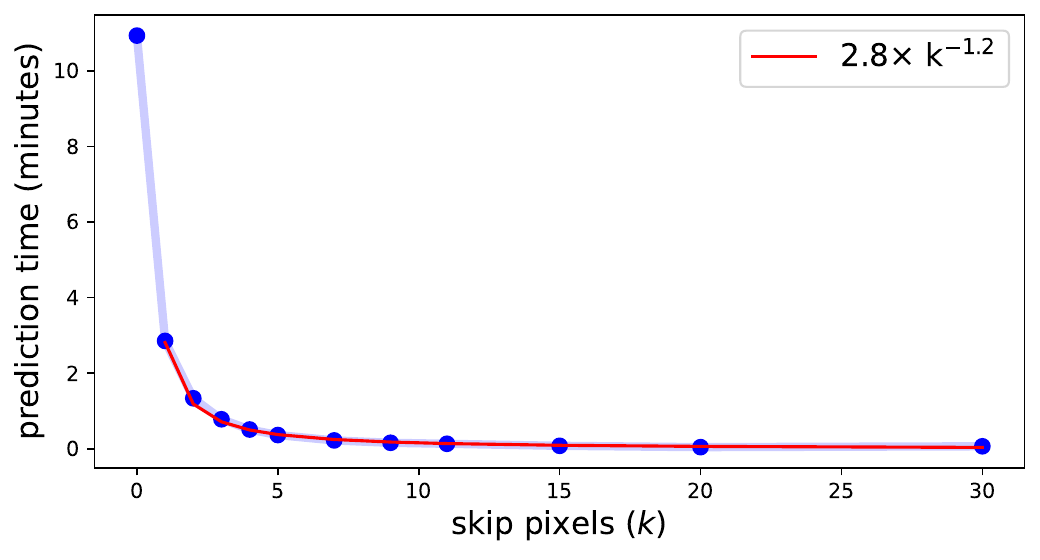}
\caption{\textbf{Prediction time for a 128x128x128 pixel-size BCDI array with a cross-shaped gap as a function of the skip pixels ($k$).} The computation time decreases as $k$ increases.}
\label{fig:skip_time}
\end{figure}

Therefore, skipping pixels can be a solution to make a fast prediction on a large BCDI array. However, skipping too many pixels also worsens the prediction accuracy. In Supplementary Fig. \ref{fig:skip_ingap}\textbf{a}, we show the ground truth detector slice in the middle of the gap. The prediction is then performed on the masked data with different skip pixel values and the results are shown in Supplementary Figs. \ref{fig:skip_ingap}\textbf{b}-\textbf{d}. Using a skip pixel value of 4 (Supplementary Fig. \ref{fig:skip_ingap}\textbf{c}) leads to a drastic decrease of the prediction time (from 11m56s down to 1m31s, a factor $\sim$8) without visible modifications. However, the drawback of using a large skip pixel value of 30 (Supplementary Fig. \ref{fig:skip_ingap}\textbf{d}) {is the occurrence of square shaped artefacts in the prediction.

\begin{figure}
\centering
\includegraphics[width=.8\textwidth,angle=0]{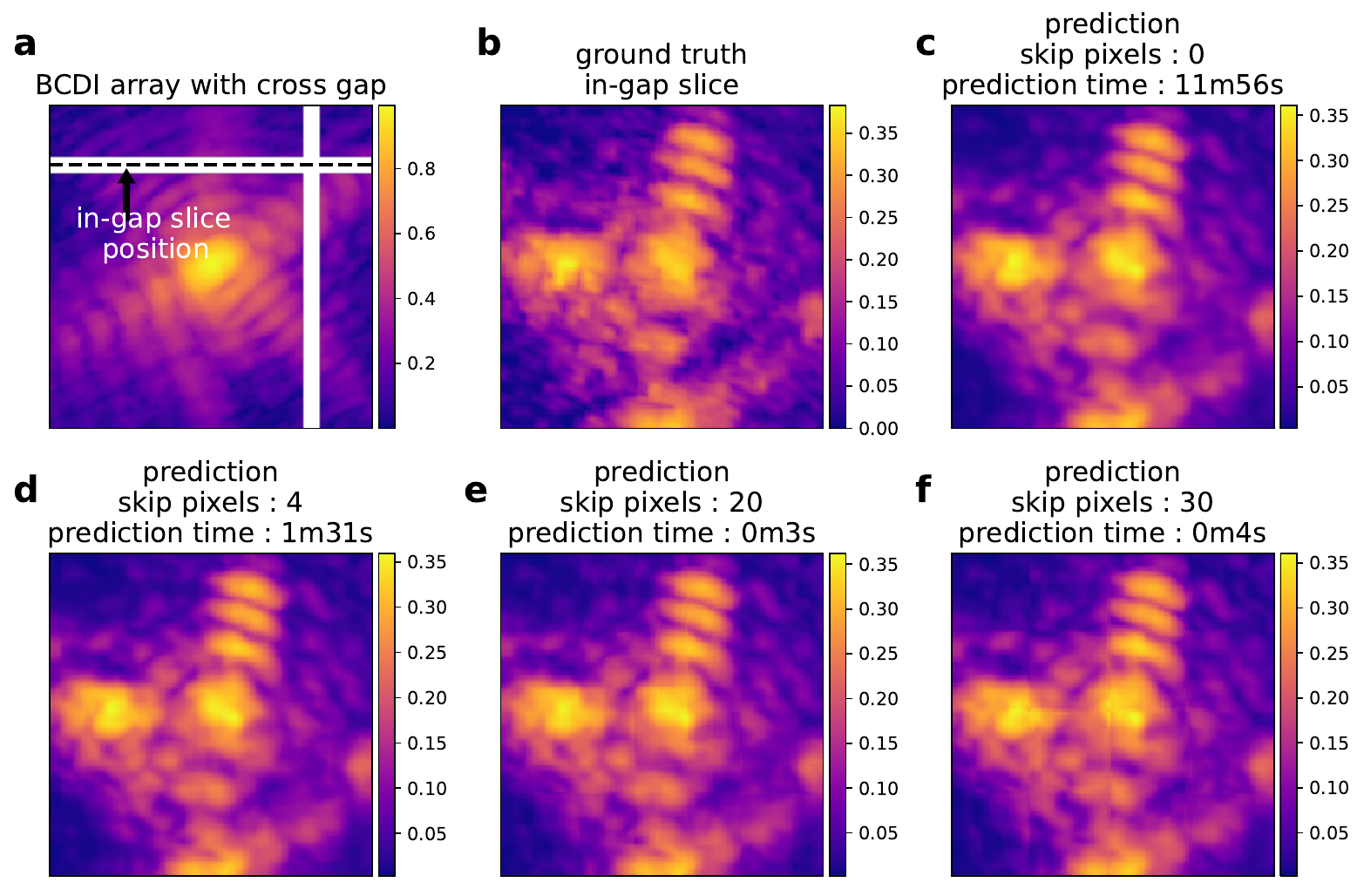}
\caption{\textbf{Predicted in-gap slice for different skip pixels.} \textbf{a} BCDI array with a cross shaped gap shown in white. The in-gap slice position is shown by a black dotted line. \textbf{b} Ground truth in-gap slice of experimental BCDI data. \textbf{c} Deep learning prediction without skipped pixels ($k$=0). \textbf{d} Prediction with $k$ = 4. \textbf{e} Prediction with $k$ = 20. \textbf{f} Prediction with $k$ = 30. In that case, square prediction artefacts are now visible.}
\label{fig:skip_ingap}
\end{figure}


\begin{figure}
\centering
\includegraphics[width=.7\textwidth,angle=0]{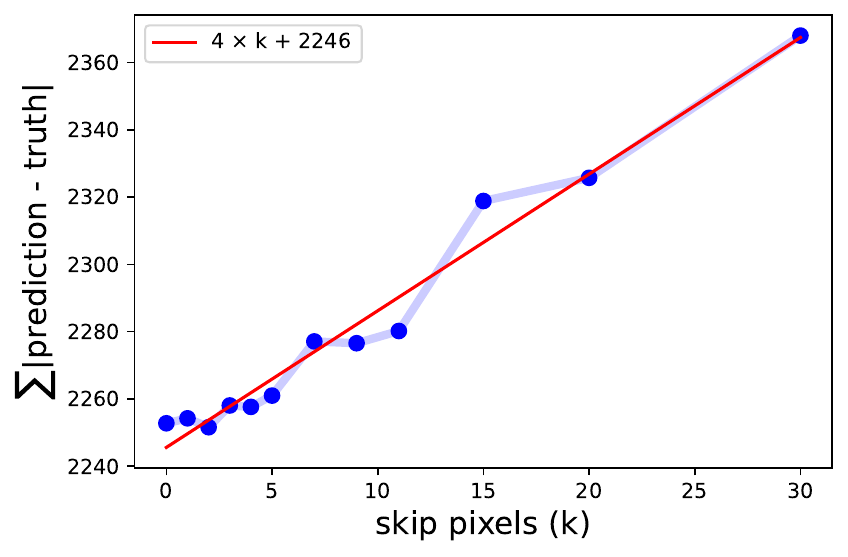}
\caption{Trend of the sum of the absolute errors between the DL prediction and the ground truth as a function of the skip pixel.}
\label{fig:skip_error}
\end{figure}

The averaged error as a function of the skip pixel value is shown in Supplementary Fig. \ref{fig:skip_error}. 
\newpage


\section{S6. Accuracy \textit{vs} oversampling}

In order to calculate the accuracy as a function of oversampling shown in Figure 6 of the main text, we simulated the same BCDI array region for several oversampling conditions. The model prediction was then computed for the whole image following the procedure illustrated in Supplementary Fig. \ref{fig:oversampling_schematic}. The model prediction is computed for different gap positions (Supplementary Figs. \ref{fig:oversampling_schematic}\textbf{a}-\textbf{b}-\textbf{c}). These predicted gaps (\textbf{d}-\textbf{e}-\textbf{f}) are then combined to produce the full predicted image (\textbf{g}). The accuracy, expressed in terms of the Pearson Correlation Coefficient is finally calculated, comparing the predicted image with the ground truth.

\begin{figure}[h]
\centering
\includegraphics[width=.67\textwidth,angle=0]{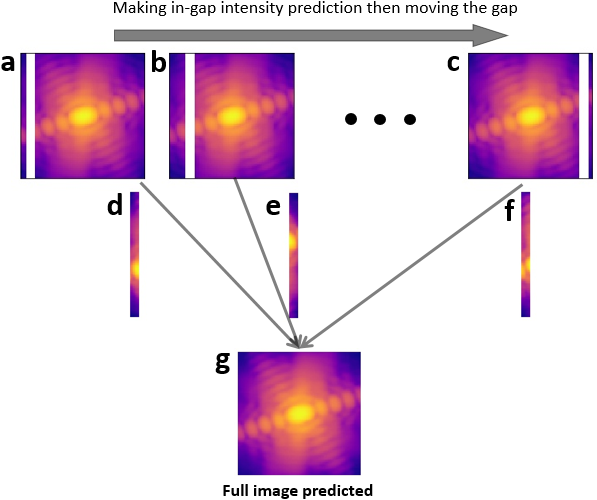}
\caption{\textbf{Schematic of the accuracy \textit{vs} oversampling calculation.} \textbf{a}-\textbf{b}-\textbf{c} Ground truth BCDI array on which a gap is applied in different positions and then predicted using our DL model. \textbf{d}-\textbf{e}-\textbf{f} Predicted gap regions. \textbf{g} Full predicted image made by combining all the predicted gaps.}
\label{fig:oversampling_schematic}
\end{figure}

Supplementary Fig. \ref{fig:oversampling_examples} shows an example of full BCDI array predictions at different oversampling ratios and for models with different gap sizes. As expected, the prediction gets worse as the oversampling becomes small and as the gap size becomes larger.

\begin{figure}
\centering
\includegraphics[width=.6\textwidth,angle=0]{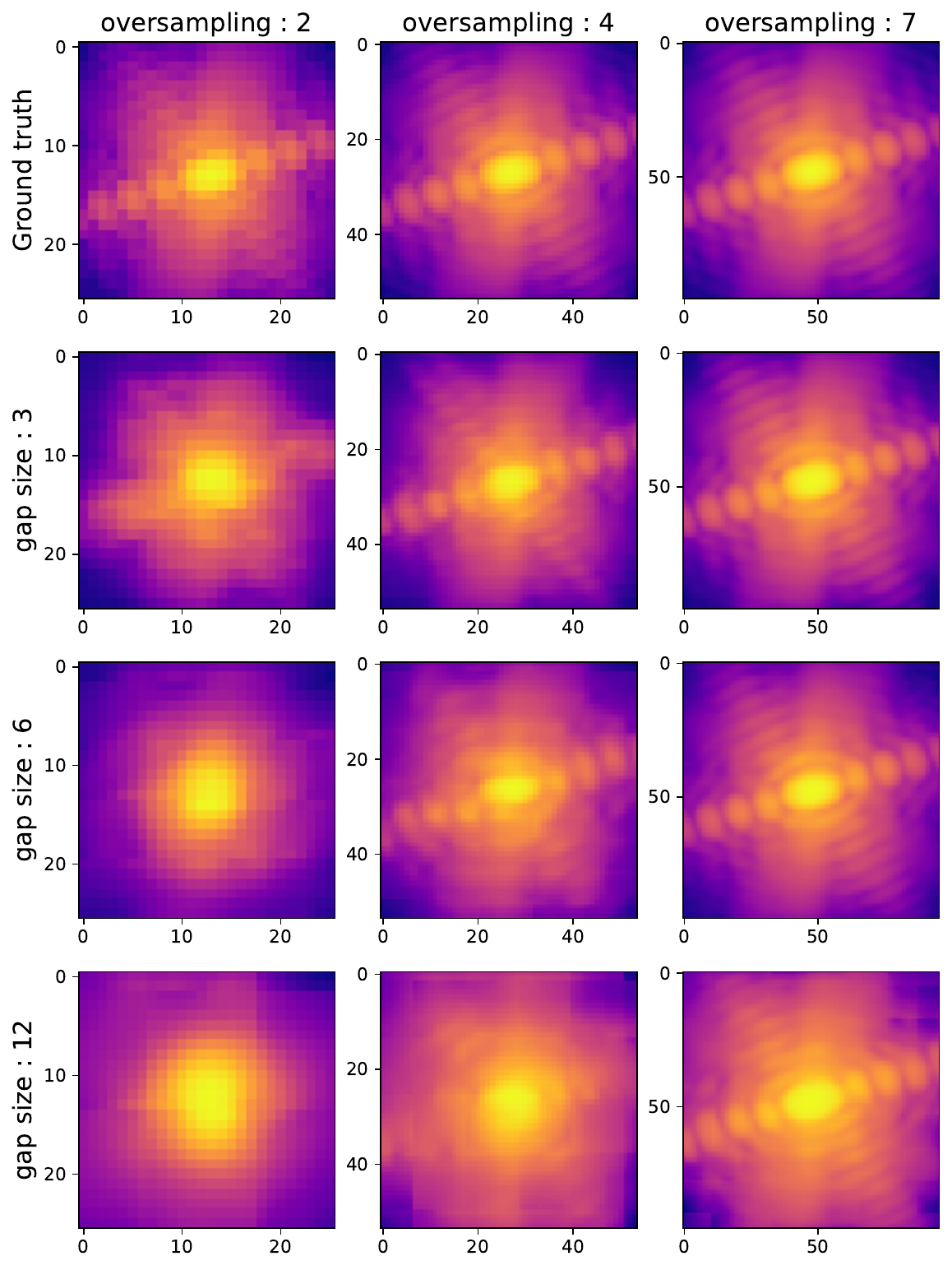}
\caption{\textbf{Full image prediction at different oversampling values and gap sizes.} 1$^{st}$ row is the ground truth BCDI arrays at different oversampling values. 2$^{nd}$-3$^{rd}$-4$^{th}$ rows show the full images predicted using DL models at different gap sizes. Best predictions occur for large oversampling and small gap size.}
\label{fig:oversampling_examples}
\end{figure}

\clearpage

\newpage

\section{S7. Real-space results}

In this section, we provide further details about the inpainting results shown in Figure 7 of the main text. The central slices of the corresponding BCDI intensities are shown in Supplementary Fig. \ref{fig:CarnisDiffractions}, respectively for the unmasked data (\textbf{a}), masked (\textbf{b}) and DL inpainted (\textbf{c}).


\begin{figure}[ht!]
\centering
\includegraphics[width=.8\textwidth,angle=0]{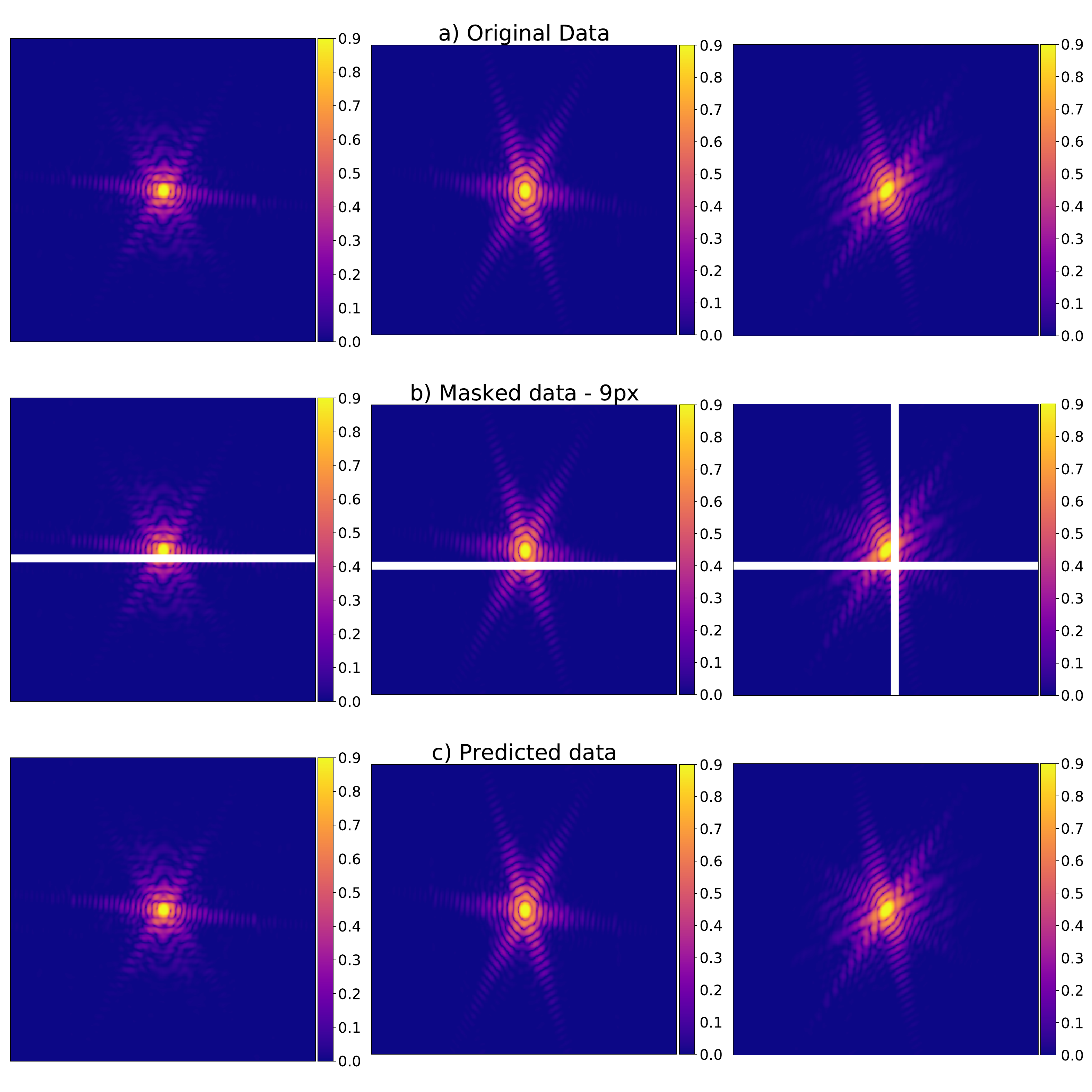}
\caption{Central slices along the three orthogonal planes of the simulated diffraction pattern treated in Ref. \cite{Carnis2019TowardsReconstructions} for the \textbf{a} Ground truth case, \textbf{b} masked with 9 pixel-wide cross-shaped gap case and \textbf{c} DL inpainted case.}
\label{fig:CarnisDiffractions}
\end{figure}


Moreover we present in Supplementary Fig. \ref{fig:strain_allgaps} the corresponding object phase for different gap widths in the simulated diffractions. As explained in the main text, the simulated phase was set to zero in the ground truth to facilitate straightforward comparisons. The particle phase variations increase with the gap size (2$^{nd}$ row of Supplementary Fig. \ref{fig:strain_allgaps}), while the DL inpainting phase (3$^{rd}$ row) remains low even with the large 12 pixel-wide gap.

\begin{figure}[h!]
\centering
\includegraphics[width=.6\textwidth,angle=0]{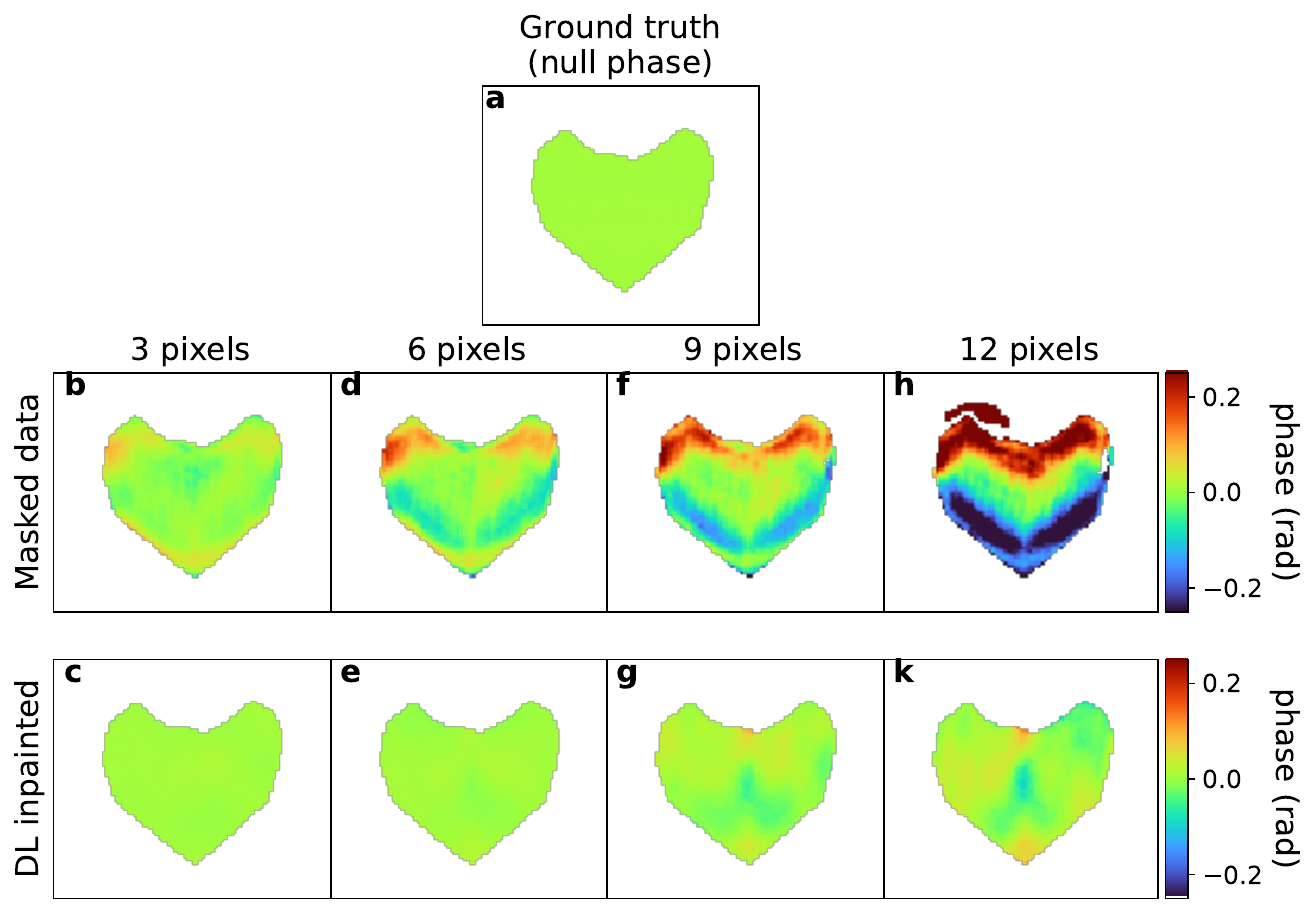}
\caption{\textbf{Object phase comparison for different gap sizes.} \textbf{a} Ground truth null phase. \textbf{b-d-f-h} Phase for masked data with different gap sizes. \textbf{c-e-g-k} Corresponding phase after DL inpainting.}
\label{fig:strain_allgaps}
\end{figure}


From the reconstructed objects at different gap widths of Supplementary Fig. \ref{fig:strain_allgaps}, we calculated the average strain per gap width shown in Supplementary Fig. \ref{fig:mean_strain}. This average strain is dropping for the reconstructed object with gap (in blue) since it introduces an asymmetry in the BCDI peak. However, our DL inpainting (in red) corrects this artefact and the average strain stays very close to zero value as it is the case for the ground truth. Variations from the average strain are represented as error bars in Supplementary Fig. \ref{fig:mean_strain}. It shows that the strain heterogeneity (visible in Supplementary Fig. \ref{fig:strain_allgaps}) is larger for the case of the masked diffraction with respect to the DL inpainted one.

\begin{figure}
\centering
\includegraphics[width=.75\textwidth,angle=0]{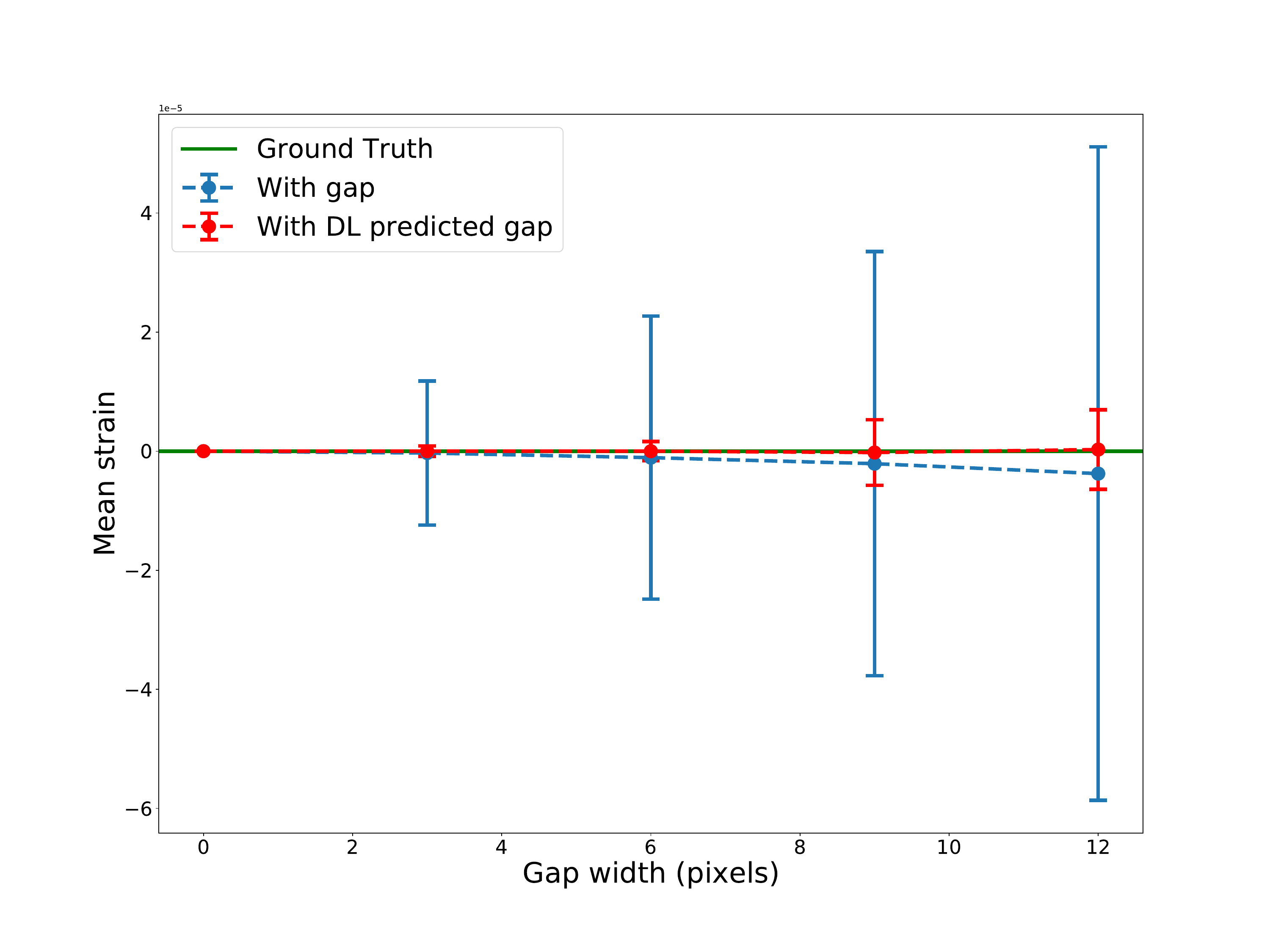}
\caption{\textbf{Average strain.} Calculation for different gap sizes for both masked and DL inpainted diffraction patterns.}
\label{fig:mean_strain}
\end{figure}

\begin{figure}[h]
\centering
  \includegraphics[width= .7\textwidth]{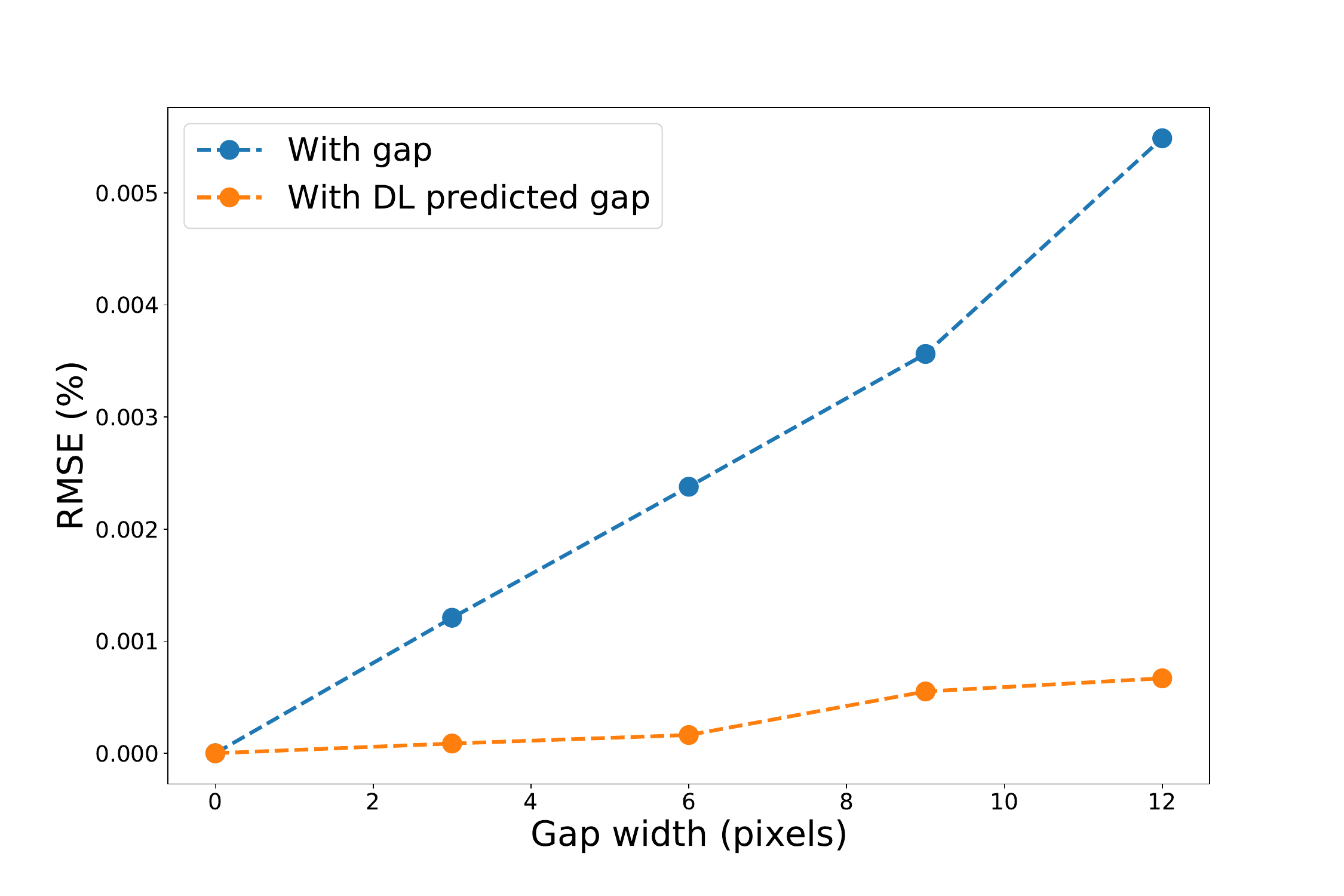}
  \caption{\textbf{Calculation of the zero-average strain root mean squared error (RMSE) \textbf{vs} the gap size} for both cases of masked and DL inpainted diffraction patterns. For all gap sizes, the DL inpainted diffraction patterns yield a smaller error.}
\label{fig:plot}
\end{figure}

\clearpage

\section{S6. Inpainting as a solution for high resolution data}

In case of large BCDI arrays used in high-resolution reconstructions, artefacts from the detector gaps can be problematic. A typical example is shown in Supplementary Fig. \ref{fig:gap_problem}\textbf{a} where the detector is relatively far from the center of the Bragg peak. However, given that the data size is 256$\times$300$\times$300 pixel-size, the number of in-gap pixels is large enough to create artefacts on the reconstructed object.


The reconstructed intensity (square modulus of the Fourier transform of the reconstructed object) is shown in Supplementary Fig. \ref{fig:high_res_recon_intensity} for several preprocessing methods. Supplementary Fig. \ref{fig:high_res_recon_intensity}\textbf{c} is the reconstructed intensity for a cropped version of the BCDI data as a reference. In \textbf{b}, a mask is used during the iterative reconstruction, leaving pixels free in the gap. However, this technique often leads to an abnormally high reconstructed in-gap intensity. In \textbf{c}, the same mask method is used but the mask is only applied near the peak streaks. In \textbf{d}, the gap is filled with 0's. Finally, in \textbf{e}, the gap is inpainted using our Deep Learning method.

\begin{figure}[h]
\centering
\includegraphics[width=.8\textwidth,angle=0]{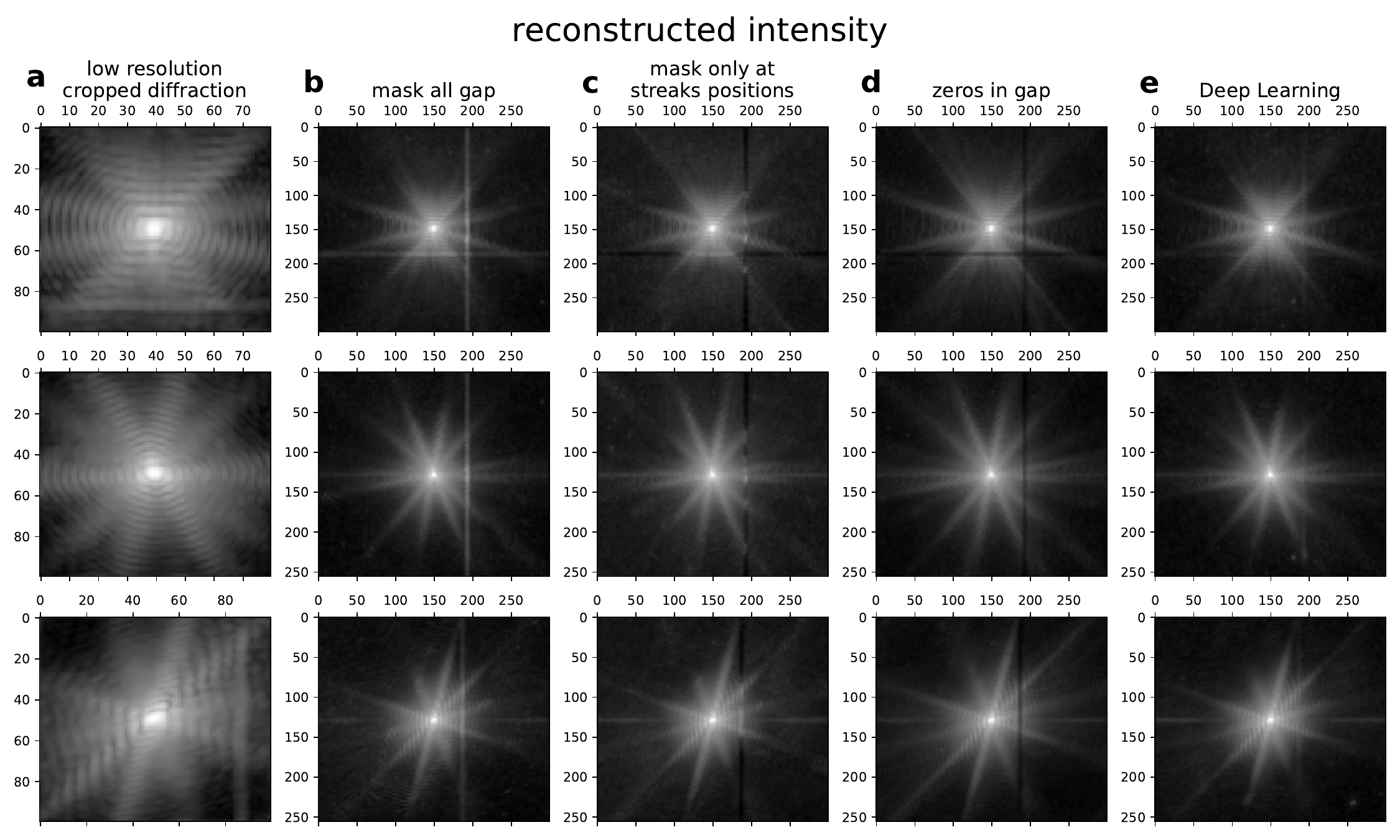}
\caption{\textbf{Reconstructed BCDI intensity comparison with different methods on experimental large BCDI array.} (in gray scale for the artefacts visibility) \textbf{a} Reference low-resolution reconstruction. \textbf{b} in-gap pixels free during the reconstruction. \textbf{c} free only in-gap pixels close to the BCDI streaks. \textbf{d} Leave 0's inside the gap. \textbf{e} Deep-Learning inpainting.}
\label{fig:high_res_recon_intensity}
\end{figure}

Supplementary Fig. \ref{fig:high_res_object_module} shows the corresponding reconstructed object modulus for the different methods. Oscillatory artefacts are clearly visible in \textbf{b} and \textbf{c} where a mask of free pixel was used during the iterative reconstruction. This is due to the additional in-gap reconstructed intensity observed for these 2 methods. When leaving the gap filled with 0's, the oscillations are less important as seen in \textbf{d} but are still visible. On the other hand, the reconstruction using Deep Learning inpainting shown in \textbf{e} does not show this artefact anymore. We emphasize that Deep Learning inpainting does not remove any high frequency information from the reconstruction, leading to the most reliable object reconstruction in this case. 

\begin{figure}
\centering
\includegraphics[width=.8\textwidth,angle=0]{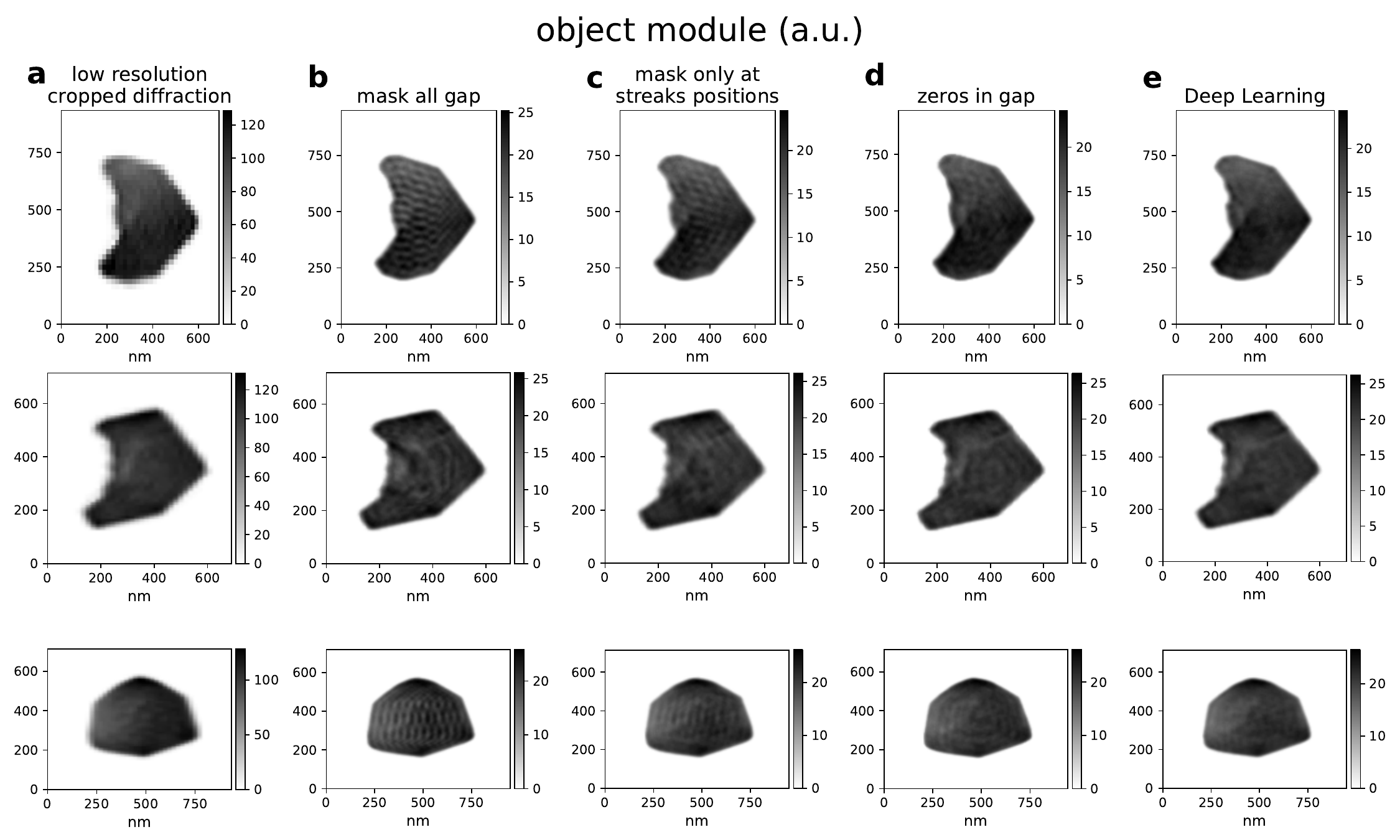}
\caption{\textbf{Reconstructed object modulus comparison with different methods on experimental large BCDI array.} Oscillatory artefacts are observed (having different magnitudes) for all methods except the Deep-Learning inpainting.}
\label{fig:high_res_object_module}
\end{figure}

Supplementary Fig. \ref{fig:high_res_object_strain} shows the reconstructed strain for the different methods. Since the strain is calculated from the phase gradient, the oscillatory artefact becomes even more important. As for the object modulus, the Deep Learning method (\textbf{e}) shows less oscillatory artefacts compared to the other methods in Supplementary Fig. \ref{fig:high_res_object_strain}\textbf{b}-\textbf{c}-\textbf{d}.

\begin{figure}
\centering
\includegraphics[width=.8\textwidth,angle=0]{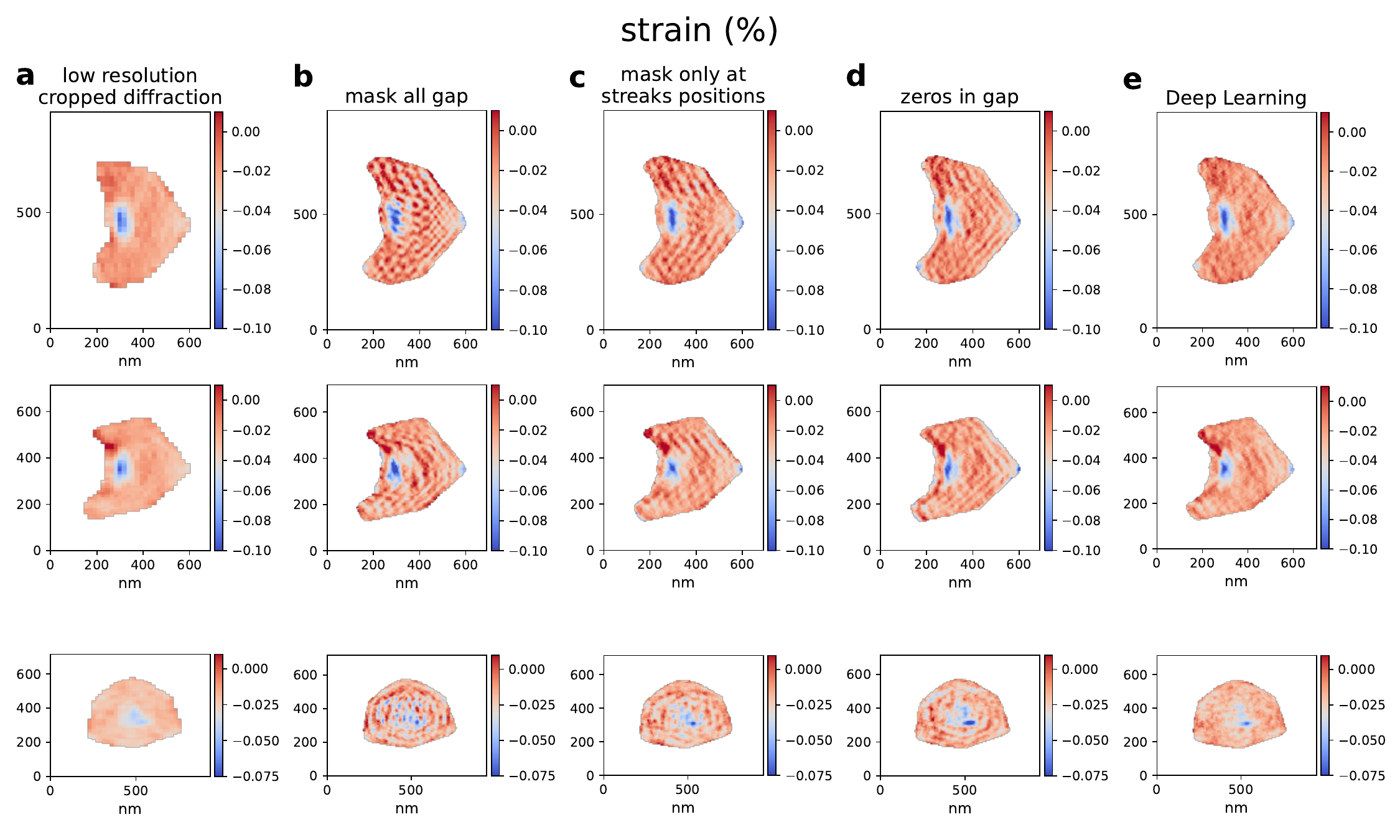}
\caption{\textbf{Reconstructed object strain comparison with different methods on experimental large BCDI array.} Oscillation artefacts are even more problematic on the strain map. Again, the Deep-Learning inpainting removes most of the oscillations.}
\label{fig:high_res_object_strain}
\end{figure}

Supplementary Fig. \ref{fig:high_res_dislocation} shows a second example of large BCDI array DL inpainting on a particle containing a dislocation. In this case as well, small oscillation artefacts on the strain maps are visible when in-gap pixels are left free during phase retrieval (\textbf{b}-\textbf{c}). Our DL inpainting manages to remove these artefacts (\textbf{e}-\textbf{f}).

\begin{figure}
\centering
\includegraphics[width=.75\textwidth,angle=0]{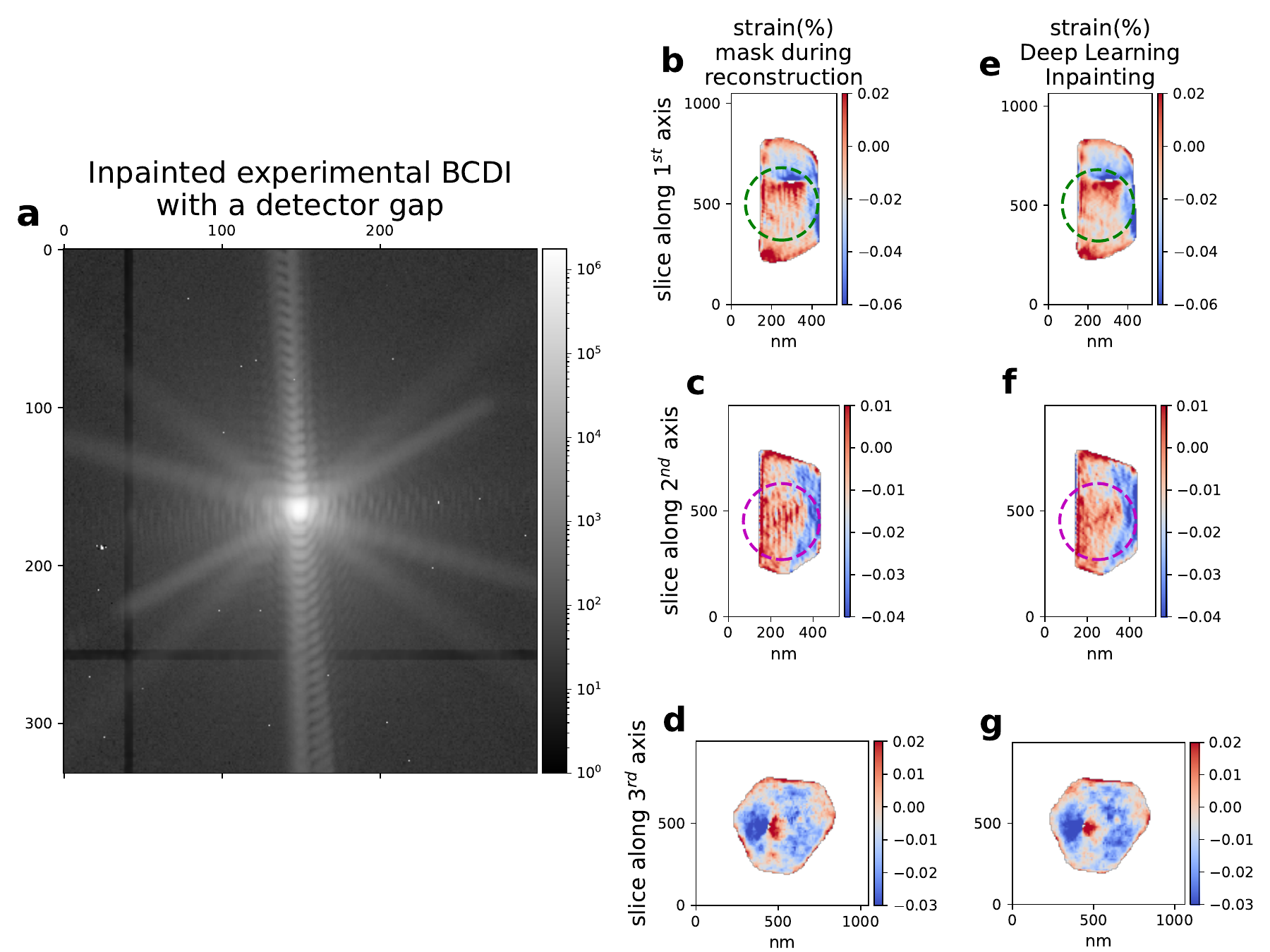}
\caption{\textbf{2$^{nd}$ example of experimental large BCDI data inpainting.} \textbf{a} Projection of the inpainted BCDI array. \textbf{b}-\textbf{c}-\textbf{d} Strain reconstruction with free in-gap pixels. \textbf{e}-\textbf{f}-\textbf{g} Strain after DL inpainting. Broken lines circles indicates region were the oscillation artefacts disappeared after DL inpainting.}
\label{fig:high_res_dislocation}
\end{figure}

\clearpage

\newpage
\bibliography{sn-bibliography}
\bibliographystyle{unsrt}

\clearpage